\documentclass{ws-ijmpa}
\usepackage[super,compress]{cite}
\usepackage{graphicx}
\begin{document}
\markboth{Authors' Names}
{Instructions for Typing Manuscripts (Paper's Title)}

%
\catchline{}{}{}{}{}
%

\title{Top-quark mass at hadron colliders}

\author{Andrea Castro\footnote{
Representing the ATLAS, CDF, CMS and D0 Collaborations. Presented at the 2013 Lepton Photon Conference, June 24-29, 2013, San Francisco, CA, USA.}}

\address{Department of Physics and Astronomy, University of Bologna, Viale Berti Pichat 6/2\\
Bologna, 40127,
Italy\\
andrea.castro@bo.infn.it}

\maketitle

\begin{history}
\received{2 April 2014}
\accepted{7 April 2014}
\published{14 August 2014}
\end{history}

\begin{abstract}
Top quarks can be produced abundantly at hadron colliders like the Tevatron at Fermilab and the Large Hadron Collider at CERN, and a variety of  measurements of top-quark properties have been gathered in the recent years from four experiments: CDF and D0 at the Tevatron and ATLAS and CMS at the Large Hadron Collider. 
In this review the most recent results on the measurement of the top-quark mass by the four  different collaborations, with various techniques and considering different topologies, are reported.
\end{abstract}

\ccode{PACS numbers: 13.85.-t,12.15Ff,14.65.Ha}


\section{Introduction}	

Since its discovery in 1995 by the CDF\cite{cdf-disc} and D0\cite{d0-disc} Collaborations, the top quark, $t$, has played a fundamental role in the Standard Model (SM) of particle physics. One of the reasons for such prominence lies in the very heavy mass of the top quark, almost 200 times that of the proton, making the top quark the heaviest among all fundamental particles. 
\par
At the hadron colliders top quarks are produced predominantly in pairs via strong interaction. At the Tevatron the production is mainly through quark-antiquark annihilation in $p\bar p$ collisions, while at the Large Hadron Collider (LHC) the pair production in $pp$ collisions is dominated by gluon-fusion diagrams. Each top quark then decays immediately into a $W$ boson and a $b$ quark, but  the different production mechanisms at the two hadron colliders does not affect the topology of the final state, $t\bar t\to W^+bW^-\bar b$, which is characterized by the way the $W$ bosons decay. When both $W$'s decay into charged leptons ($e$ or $\mu$) plus the corresponding neutrinos, the final state called {\it dilepton} has a small branching ratio ($BR\approx 5\%$); when only one leptonic decay of the $W$'s is occuring the final state, named {\it lepton+jets}, has a larger $BR\approx 30\%$; finally in case of no leptonic $W$ decays, the branching ratio in the so-called {\it all-jets} (or {\it all-hadronic}) channel is the highest: $BR\approx 46\%$. Final states where $W$'s decay into a $\tau$ and its neutrino are usually treated separately.
\par
The difference  in the center-of-mass energy, $\sqrt{s}$, of the two colliders translates into a large difference in the $t\bar t$ production rate. At the Tevatron, with $\sqrt{s}=1.96$ TeV,  the production cross section is $\sigma_{t\bar t}\approx 7$ pb\cite{top-xsec}, assuming a top-quark mass $m_t=173.5$ GeV\cite{world-ave-ref}, while at the LHC the cross section becomes\cite{top-xsec} $\sigma_{t\bar t}\approx 170$ pb when running at $\sqrt{s}=7$ TeV, or  $\sigma_{t\bar t}\approx 250$ pb when running at $\sqrt{s}=8$ TeV.
\par
The status of top-quark mass measurements at the Tevatron before the LHC turn-on was discussed in Ref.~\refcite{tev-review}, while Ref.~\refcite{lhc-review} contains a recent comprehensive review of top-quark physics at the LHC.

\par
In these proceedings we will discuss the most recent measurements of the top-quark mass performed at the hadron colliders by four experiments: CMS and D0 at the Tevatron, with up to $8.7$ fb$^{-1}$ of integrated luminosity, and ATLAS and CMS at the LHC, with up to $5.0$ fb$^{-1}$ of integrated luminosity. We will show how the data accumulated over the years and the application of different strategies and methods made the top quark the quark whose mass is measured with the highest precision, of about 0.5\%.

\section{Motivations for the Top-Quark Mass Measurement}
The measurement of the top-quark mass is an important part of the physics program at the Tevatron and at the LHC.
The top quark is peculiar among all quarks because it decays quickly, well before hadronizing, so we can measure its mass, $M_t$, which is  a free parameter of the SM,  directly from the observation of its decay products. Such a  measurement has been strongly pursued in the past 18 years with an accuracy improving from about 4 GeV at the time of the top-quark discovery, to less than a GeV of the current measurements. 
\par
\begin{figure}[hb]
\begin{tabular}{cc}
\includegraphics[width=6.8cm]{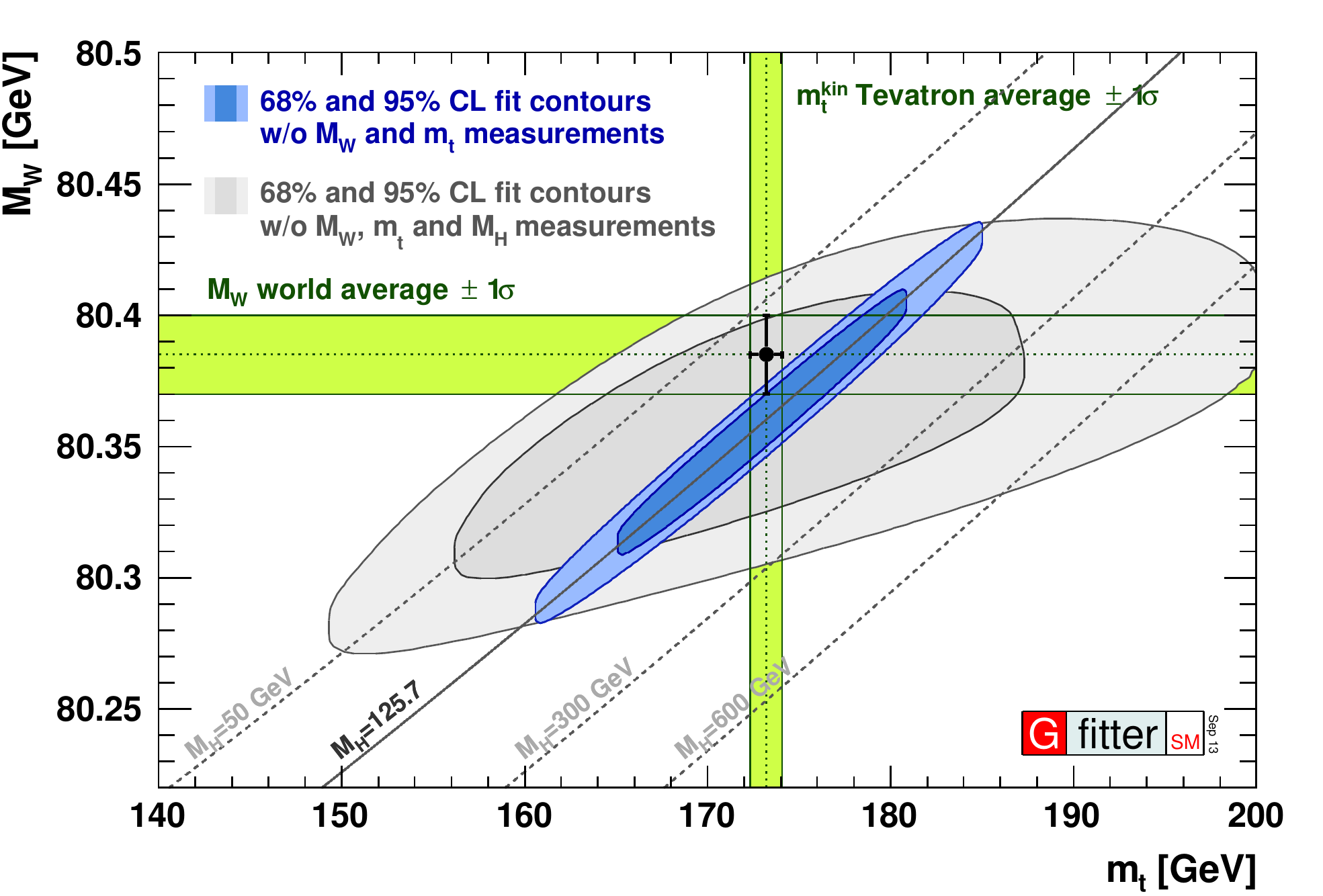}
\includegraphics[width=6.8cm]{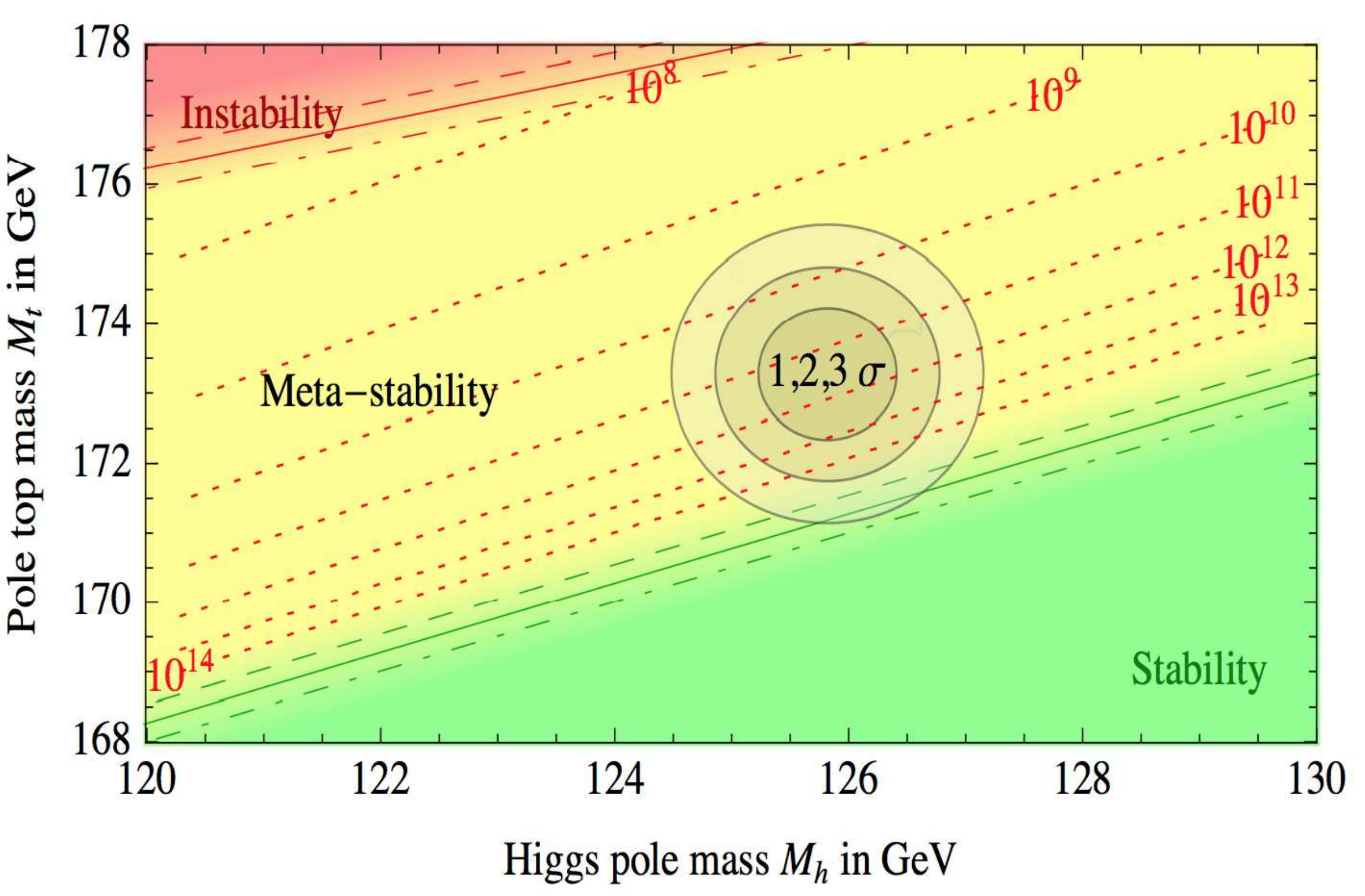}
\end{tabular}
\caption{(Left) Global fit of measurements in the SM for the mass of $W$, $t$ and $H$. (Right) Regions of stability for the SM as a function of the top-quark and Higgs-boson masses.}
\label{fit-and-vacuum}
\end{figure}

Another reason for measuring $M_t$ lies in the fact that top quarks participate to quantum loop radiative corrections to the $W$-boson mass. Before the discovery of the Higgs boson, $H$, these corrections were quite important because they provided a way to constrain the mass of the Higgs boson.  Now, after the Higgs boson discovery by ATLAS\cite{atlas-higgs-ref} and CMS\cite{cms-higgs-ref}, these masses are critical inputs to global electroweak fits\cite{sm-global-fit} which assess  the self-consistency within the SM, as shown in Fig.~\ref{fit-and-vacuum} (left).
\par
The huge mass of the top quark puts it close to the scale of the electroweak symmetry breaking. For this reason the top quark might play a special role in it or in instances of new physics like in topcolor models\cite{topcolor-ref1, topcolor-ref2} for electroweak dynamical breaking\cite{ewk-break-ref}.
\par
Finally, the top-quark mass is related together with the Higgs-boson mass to the vacuum stability\cite{vacuum-stability} of the SM. In fact the value of the Higgs-boson mass, about 125 GeV, measured by ATLAS\cite{atlas-higgs-ref} and CMS\cite{cms-higgs-ref}, is crucial because it is quite close to the minimum value that ensures absolute vacuum stability within the SM, see Fig.~\ref{fit-and-vacuum} (right).

\section{Physics Signatures}
The multipurpose detectors, CDF and D0 at the Tevatron and ATLAS and CMS at the LHC, have  some common features. They are essentially spectrometers with: tracking capability for measuring the transverse momentum, $p_T$, of charged particles, complemented by high-precision (microvertex) tracking;  electromagnetic calorimetry to measure electrons and photons;  hadronic calorimetry to measure jets;  muon systems to identify and measure muons.
\par
The physics signatures provided by these detectors and of interest for the top-quark mass measurement are then: high-$p_T$ isolated leptons ($e$ or $\mu$); high-$p_T$ jets, some of which can be identified as initiated from a $b$ quark (i.e. $b$-jets); missing transverse energy, $E_T^{miss}$, corresponding to the transverse momentum imbalance associated to neutrinos. It is worth mentioning that jets, as measured from the energy deposit in the calorimeters, need corrections in order to derive the energy of the corresponding partons. Different algorithms have been introduced (for instance, cone clustering of different types, or the CMS particle-flow algorithm which uses also the information from the tracker) to reduce the amount of correction needed and the related systematic uncertainties, and to improve the resolution. Even after these corrections, the absolute value of the jet energy is not perfectly known and the corresponding so-called jet energy scale (JES) is still known with some uncertainty and treated with an overall factor with respect to the nominal calibration.

\section{Measuring the Top-Quark Mass}
The measurement of $M_t$ has been performed at the Tevatron and LHC with different techniques having complementary and competing features. 
All of them start however from the reconstruction of the $p\bar p ~ (pp)\to t\bar t\to WbW\bar b$ final state from the kinematic variables of the leptons/jets associated to the final state.
\par
There are several issues one has to deal with. First, the choice of the final state topology and the necessary event selection. Then, the mapping of the leptons/jets reconstructed in the event to the leptons/partons of the expected final state, and this carries along ambiguities and combinatorial issues. Another important ambiguity is related to detector modeling issues like the uncertainty on the energy calibration of the detector. Finally, there are unknown quantities, like the longitudinal component of the neutrino momentum, $p_z^\nu$,  or the sharing of the $E_T^{miss}$ between multiple neutrinos. For these cases, corresponding to the dilepton channel,  the kinematics of the final state is underconstrained.

\subsection{Methods for measuring $M_t$}
Once the final state is reconstructed, there are several methods which have been applied so far at the Tevatron and LHC for the measurement of the top-quark mass; the most common are:
\begin{itemize}
\item the {\em template} method;
\item the {\em ideogram} method;
\item the {\em matrix element} method;
\item and the {\em matrix-weighting} (or {\em neutrino-weighting} or {\em kinematic analysis}) techniques;
\end{itemize}
which will be described briefly below.

\subsubsection{Template and ideogram methods}
\begin{figure}[hb]
\begin{tabular}{cc}
\includegraphics[width=6.8cm]{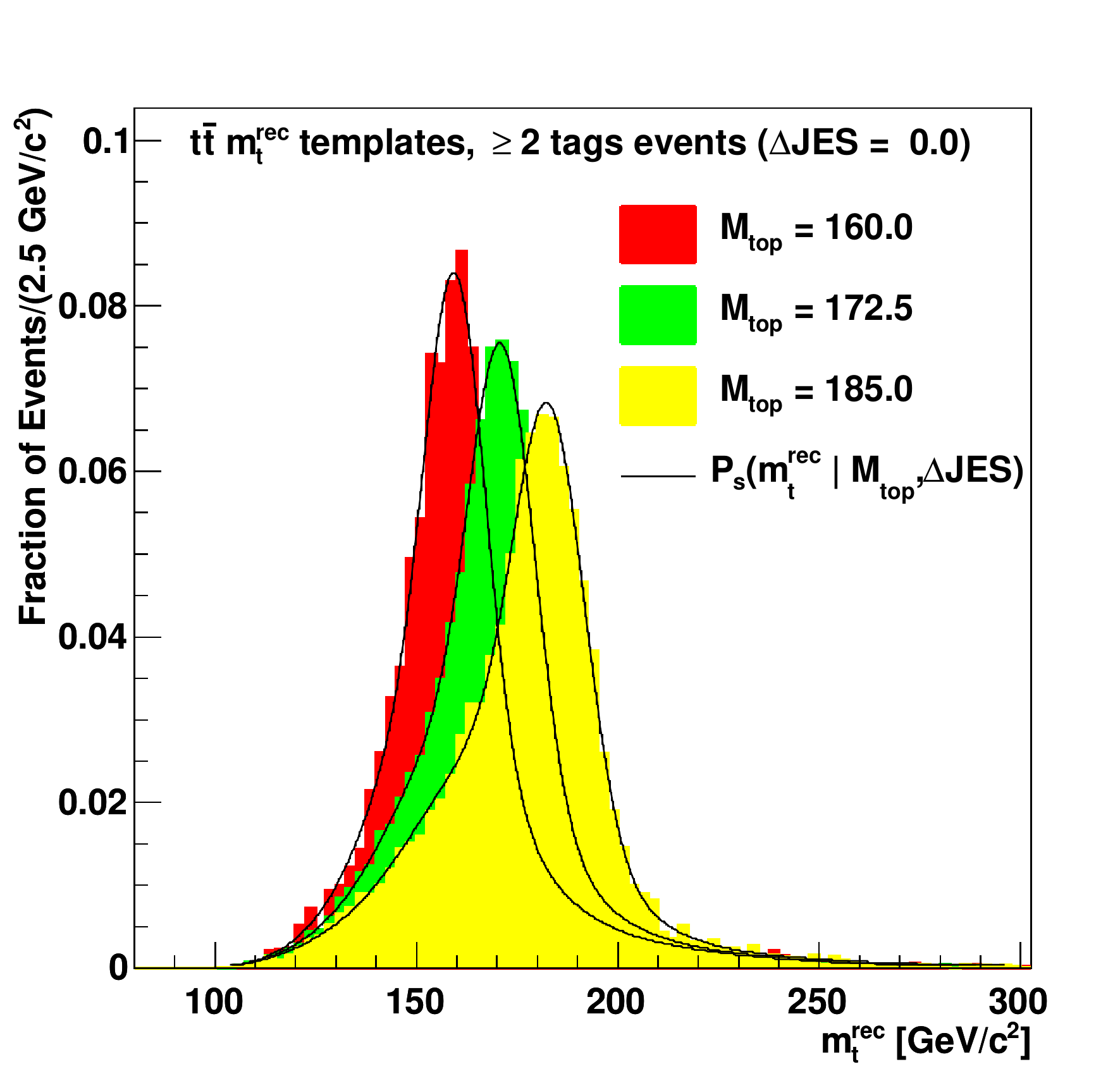}
\includegraphics[width=6.8cm]{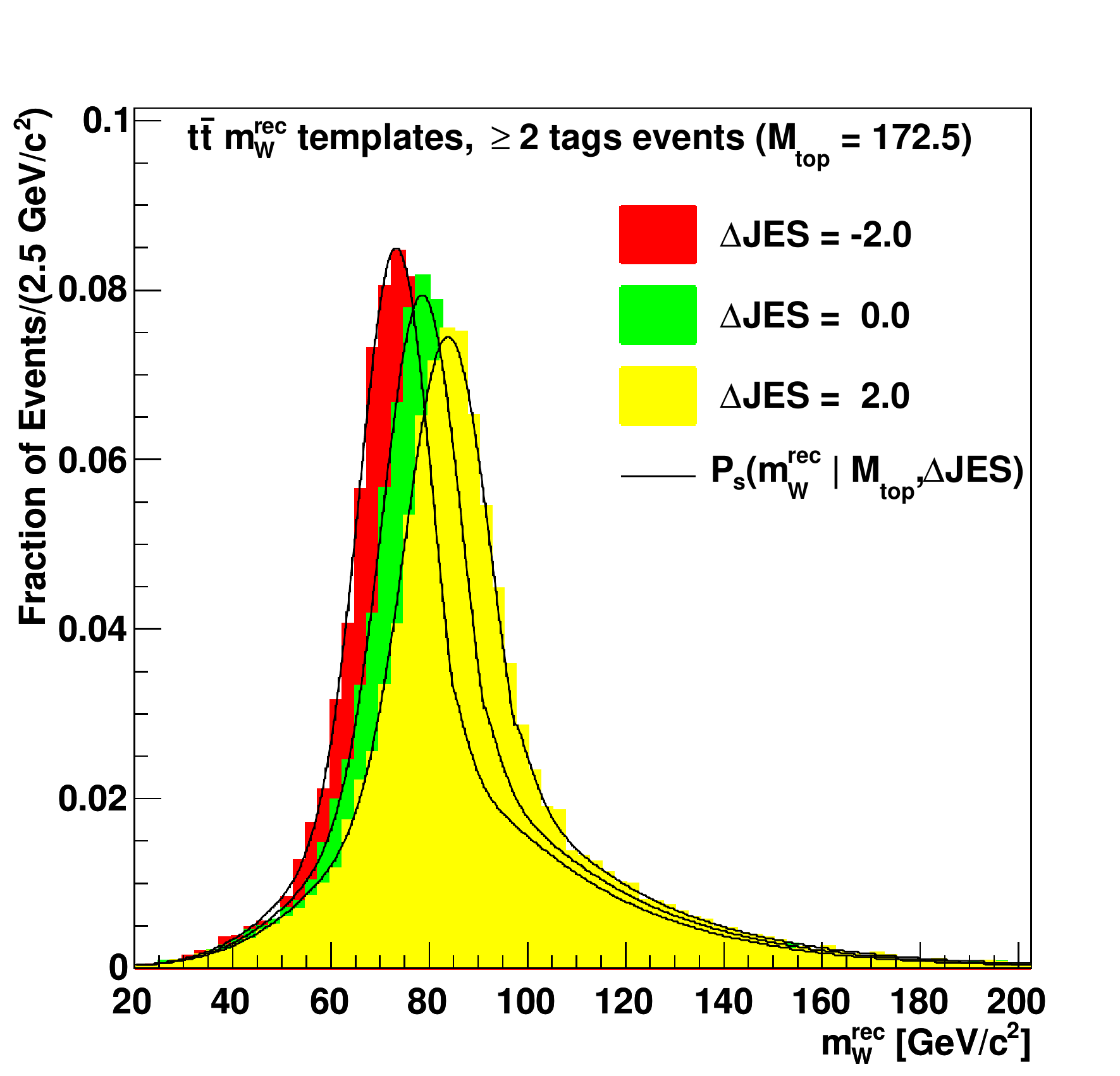}
\end{tabular}
\caption{CDF simulations for the all-jets channel. (Left) Distribution of the reconstructed top-quark mass for different values of $M_t$. (Right)  Distribution of the reconstructed $W$-boson mass for different values of JES shifts defined in terms of energy uncertainties.}
\label{templates-mass}
\end{figure}

The template method is based on distributions of variables sensitive to $M_t$. The typical choice is the reconstructed top-quark mass from a $\chi^2$ fit to the $WbWb$ hypothesis for the final state. The combination which yields the smallest $\chi^2$ is usually chosen to represent the reconstructed top-quark mass, while the $W$ boson mass, $M_W$, is constrained to its measured value. Distributions (templates) are then derived for Monte Carlo (MC) generated events, assuming different values of $M_t$, and analytical functions representing their probability density are parametrized as a function of $M_t$. A likelihood is then computed based on these distributions or functions. The inclusion of $M_W$ templates depending explicitly on JES shifts from the nominal value, allows for an {\em in-situ} calibration of the JES. It is possible to consider also constraints on the JES of the $b$-jets.  An example of the behavior of templates as a function of $M_t$ and the JES shifts  can be seen in Fig.~\ref{templates-mass}, which refers to simulations of all-jets events at CDF.
\par
In general this method is quite simple and fast, but affected by statistical uncertainties which are typically larger with respect to other methods.
\par
A generalization of the template method is the {\em kernel density estimate} which is a parametric $N$-dimensional version.
\par
The ideogram method\cite{ideogram-ref} is a modification of the template method which accounts for the $M_t$ resolution on an event-by-event basis. The method starts from the kinematical reconstruction of the $WbWb$ final state and then computes an event likelihood as a function of $M_t$, convoluting Breit-Wigner (or similar) distributions with experimental resolutions.

\subsubsection{Matrix element method}
The matrix element method, pionereed by the D0 experiment\cite{matrix-ref}, computes the probability to obtain the observed set ${\bf x}$ of variables given an assumed top-quark mass $M_t$ and a generated set  ${\bf y}$ of variables. In this case full information on the event is considered and compared to the theoretical expectation derived from the matrix element, the parton distribution functions (PDF) $f(q)$, and the application of the appropriate transfer functions $W({\bf x},{\bf y})$. A probability is defined to have an event compatible with the $t\bar t$ hypothesis, given an assumed value of $M_t$. Such a probability is calculated as:
\begin{equation}
P(t\bar t,M_t)\propto \int\sum_{\rm flavors}dq_1dq_2\frac{d\sigma(p\bar p (pp)\to t\bar t\to {\bf y}; M_t)}{d{\bf y}}f(q_1)f(q_2)W({\bf x},{\bf y})d{\bf y}.
\end{equation}
An event probability is then defined in terms of $P(t\bar t,M_t)$ plus a background probability. Finally a total likelihood is computed and maximized to obtain the best $M_t$ value.

\subsubsection{Matrix-weighting techniques}
These methods (named also neutrino-weighting or kinematic analysis) are usually applied in the case of the dilepton channel where a given $M_t$ value is used to constrain the $t\bar t$ system, inferring the neutrino momenta from the $E_T^{miss}$, and assuming values for the unobserved quantities like $p_z^\nu$. Weights are assigned to the possible solutions and templates are built from these weights.

\subsection{Systematic uncertainties}
Large samples of $t\bar t$ events have been collected at the Tevatron, and even larger at the LHC. This means that, in general, the statistical uncertainties are small and the measurements are instead dominated by systematic uncertainties.
\par
There are several sources of systematic uncertainty which need to be considered. Among them, one of the most relevant is the effect due to the imperfect knowledge of the JES for generic jets or for $b$-flavored jets (bJES) which translate into overall scale factors (JSF or bJSF) which might be different from 1. Other important sources of systematic uncertainty are related to the signal modeling, and are evaluated by using different MC generators, hadronization models, color-reconnection schemes, varying the amount of underlying events and of initial/final state radiation (ISR/FSR), or choosing different PDFs. In addition there are uncertainties related  to the background modeling or the lepton energy/momentum determination. Finally there are uncertainties associated to specific features of the method applied and the size of the MC samples used.

\section{Top-Quark Mass Measurements at the Tevatron}
At the Tevatron a large number of measurements of the top-quark mass have been performed over the years by CDF and D0, based on integrated luminosities up to 8.7 fb$^{-1}$. We present here only the most accurate results for each experiment in the most relevant $t\bar t$  channels.
\subsection{Lepton+jets channel}
\begin{figure}[hb]
\begin{tabular}{cccc}
   \begin{minipage}{0.4\textwidth}
      \includegraphics[width=6.8cm]{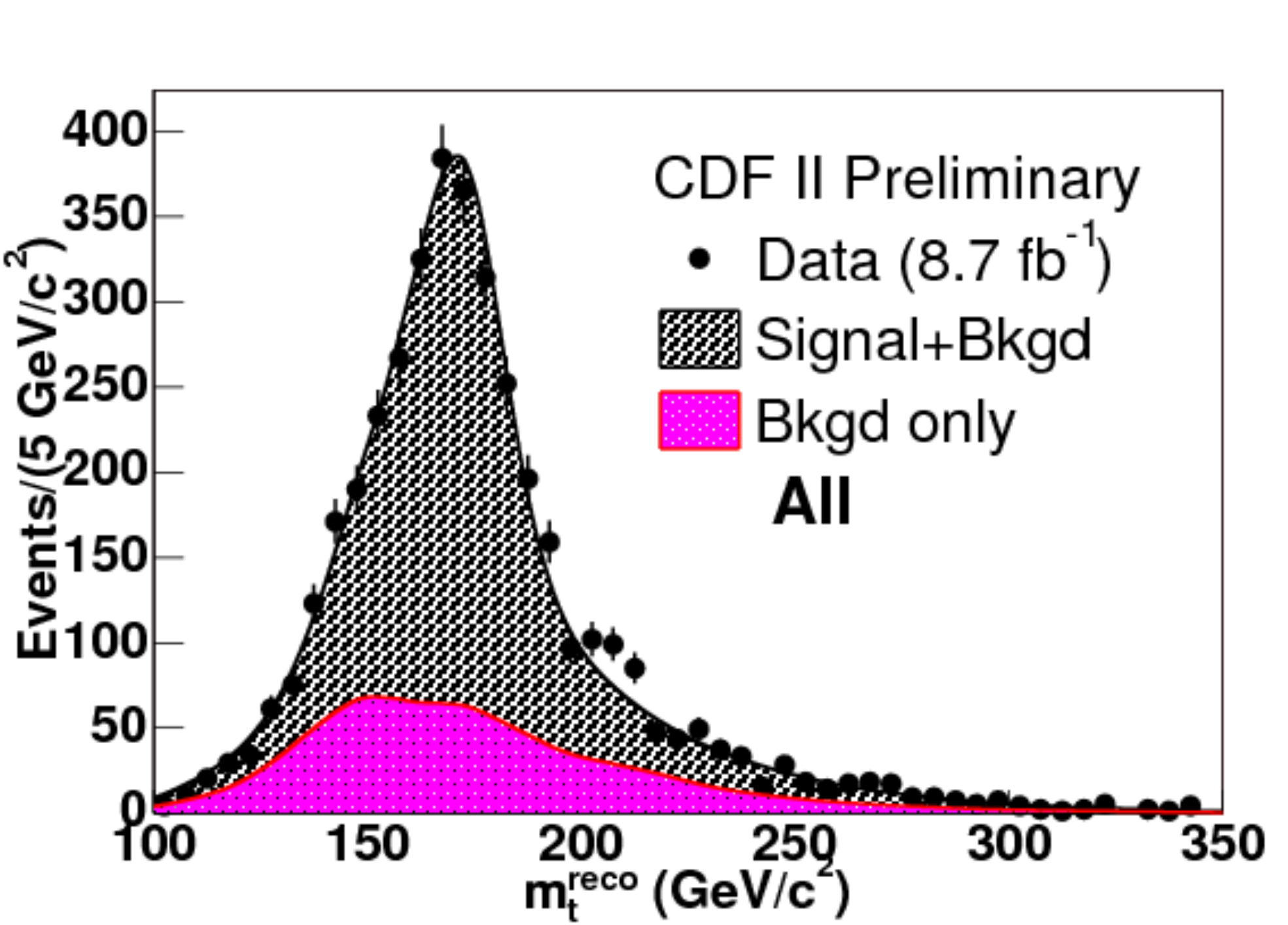}
   \end{minipage}
  & ~~~ & ~~~ &
   \begin{minipage}{0.4\textwidth} 
      \includegraphics[width=6.8cm]{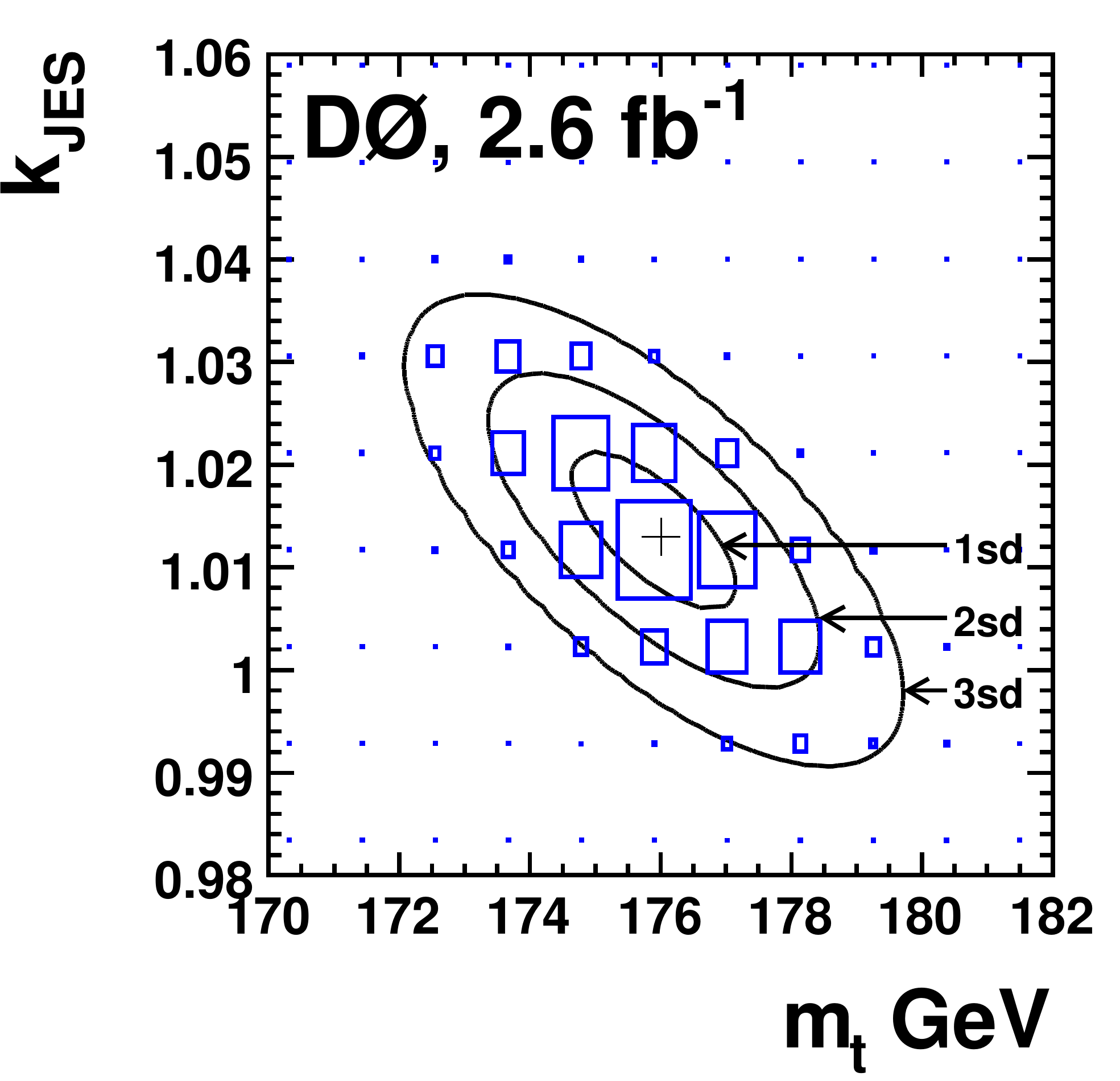}
   \end{minipage}
\end{tabular}
\caption{(Left) Reconstructed top-quark mass for CDF lepton+jets data (8.7 fb$^{-1}$). (Right) Fitted contours of equal probability for D0 lepton+jets data (2.6 fb$^{-1}$), in terms of $M_t$ and JSF (named $k_{\rm JES}$ in the plot).}
\label{cdf-d0-lj}
\end{figure}

The lepton+jets channel is expected to be the golden one, providing the most accurate measurements with respect to other channels. 
Recurring to the kernel density estimation method with {\em in-situ} JES calibration, CDF measures\cite{cdf-lj-ref} a top-quark mass $M_t=172.85\pm 0.52({\rm stat.})\pm 0.49({\rm JES})\pm 0.84({\rm syst.})$ GeV, with a total uncertainty of 1.1 GeV. The systematic uncertainty is dominated by contributions due to the generator modeling (0.56 GeV) and the residual JES (0.52 GeV). The expected and observed distributions of the reconstructed top-quark mass are shown in Fig.~\ref{cdf-d0-lj} (left).
\par
D0 applies instead the {\em matrix element} method including an {\em in-situ} calibration of JES, measuring\cite{d0-lj-ref} $M_t=174.94\pm 0.83({\rm stat.})\pm 0.78({\rm JES})\pm 0.96({\rm syst.})$ GeV, with a total uncertainty amounting to 1.5 GeV. The systematic uncertainty is dominated by contributions due to the generator modeling (0.58 GeV) and the jet energy resolution (JER) knowledge (0.32 GeV).  The fitted Gaussian contours of equal probability for the two-dimensional likelihoods as a function of $M_t$ and JSF are shown in Fig.~\ref{cdf-d0-lj} (right).

\subsection{Dilepton or $E_T^{miss}$+jets channels}
\begin{figure}[hb]
\begin{tabular}{cccc}
   \begin{minipage}{0.4\textwidth}
      \includegraphics[width=6.8cm]{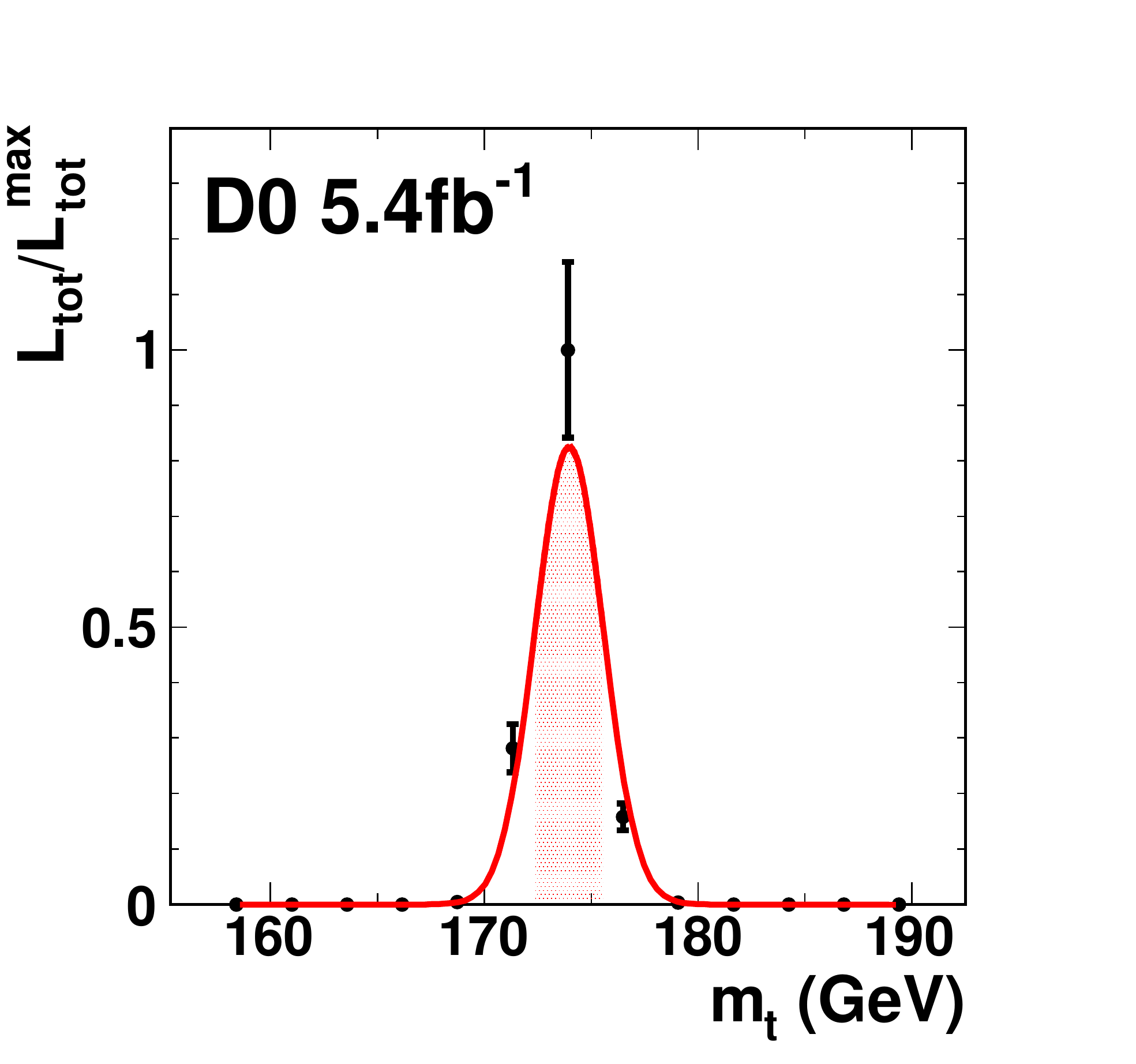}
   \end{minipage}
  & ~~~ & ~~~ &
   \begin{minipage}{0.4\textwidth} 
      \includegraphics[width=6.8cm]{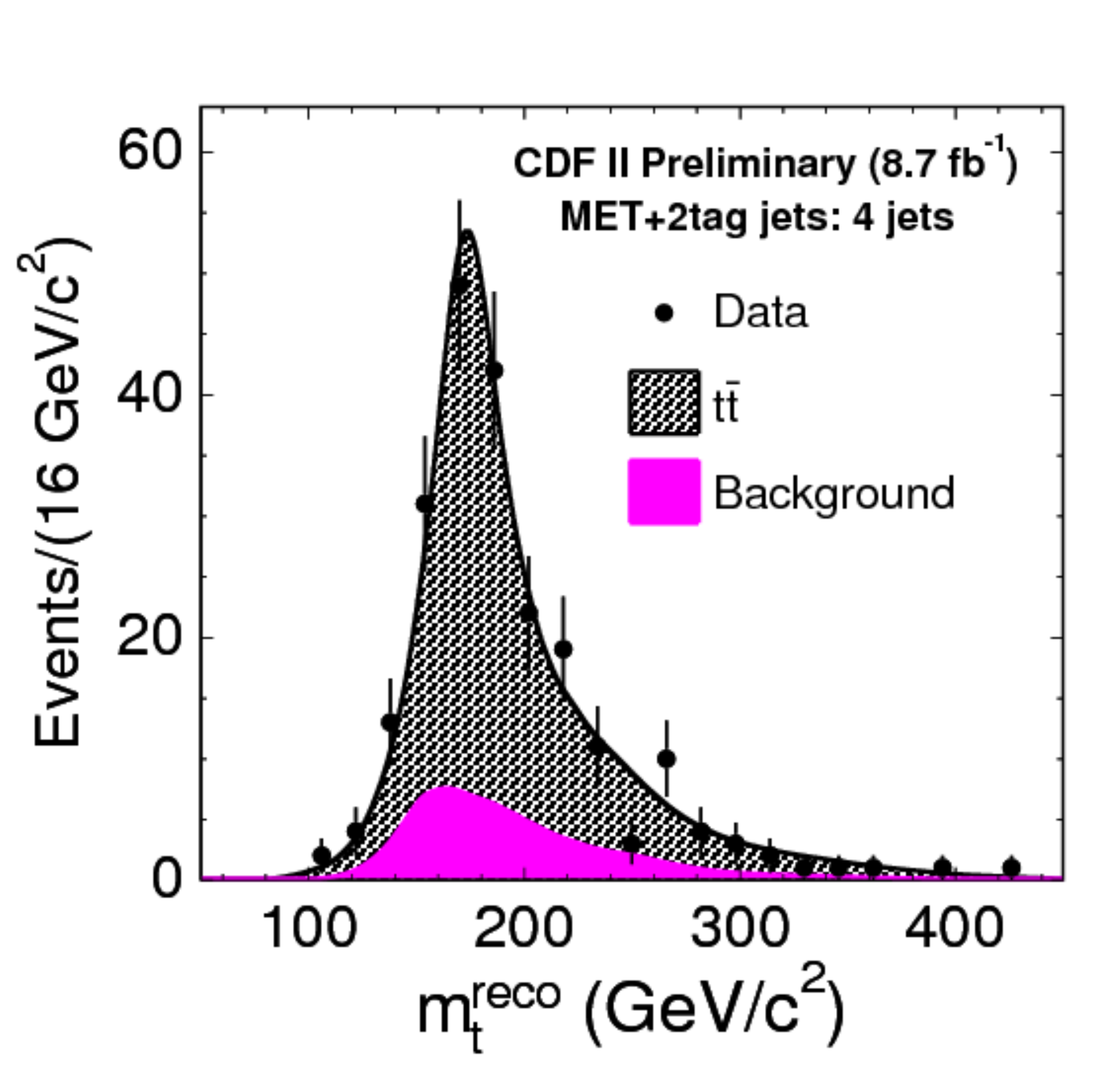}
   \end{minipage}
\end{tabular}
\caption{(Left) Calibrated and normalized likelihood for D0 dilepton data (5.4 fb$^{-1}$) as a function of the top-quark mass.
(Right) Reconstructed top-quark mass for CDF $E_T^{miss}$+jets data (8.7 fb$^{-1}$).}
\label{cdf-d0-dil}
\end{figure}

A measurement in the  dilepton channel is obtained by D0 combining the matrix element and neutrino-weighting techniques. The measurement\cite{d0-dil-ref} gives  a value $M_t=173.9\pm 1.9({\rm stat.})\pm 1.6({\rm syst.})$ GeV, with a total uncertainty of 2.5 GeV. The systematic uncertainty is dominated by contributions due to the generator modeling (0.6 GeV) and the JES (0.9 GeV). The measurement is represented in  Fig.~\ref{cdf-d0-dil} (left).
\par
CDF considers also the contributions from $W\to \tau \nu$ decays by looking at events with jets plus large $E_T^{miss}$, obtaining a precision better than using the regular dilepton topology.  The measurement\cite{cdf-dil-ref} is based on a kernel density estimation plus {\em in-situ} JES calibration and gives  a value $M_t=173.9\pm 1.3({\rm stat.})\pm 1.1({\rm JES})\pm 0.9({\rm syst.})$ GeV, with a total uncertainty of 1.9 GeV. The systematic uncertainty is dominated by contributions due to the generator modeling (0.4 GeV) and the JES (0.4 GeV).  The expected and observed distribution of the reconstructed top-quark mass for events with 2 b-tagged jets are shown in Fig.~\ref{cdf-d0-dil} (right).

\subsection{All-jets channel}
\begin{figure}[hb]
\centerline{\includegraphics[width=6.8cm]{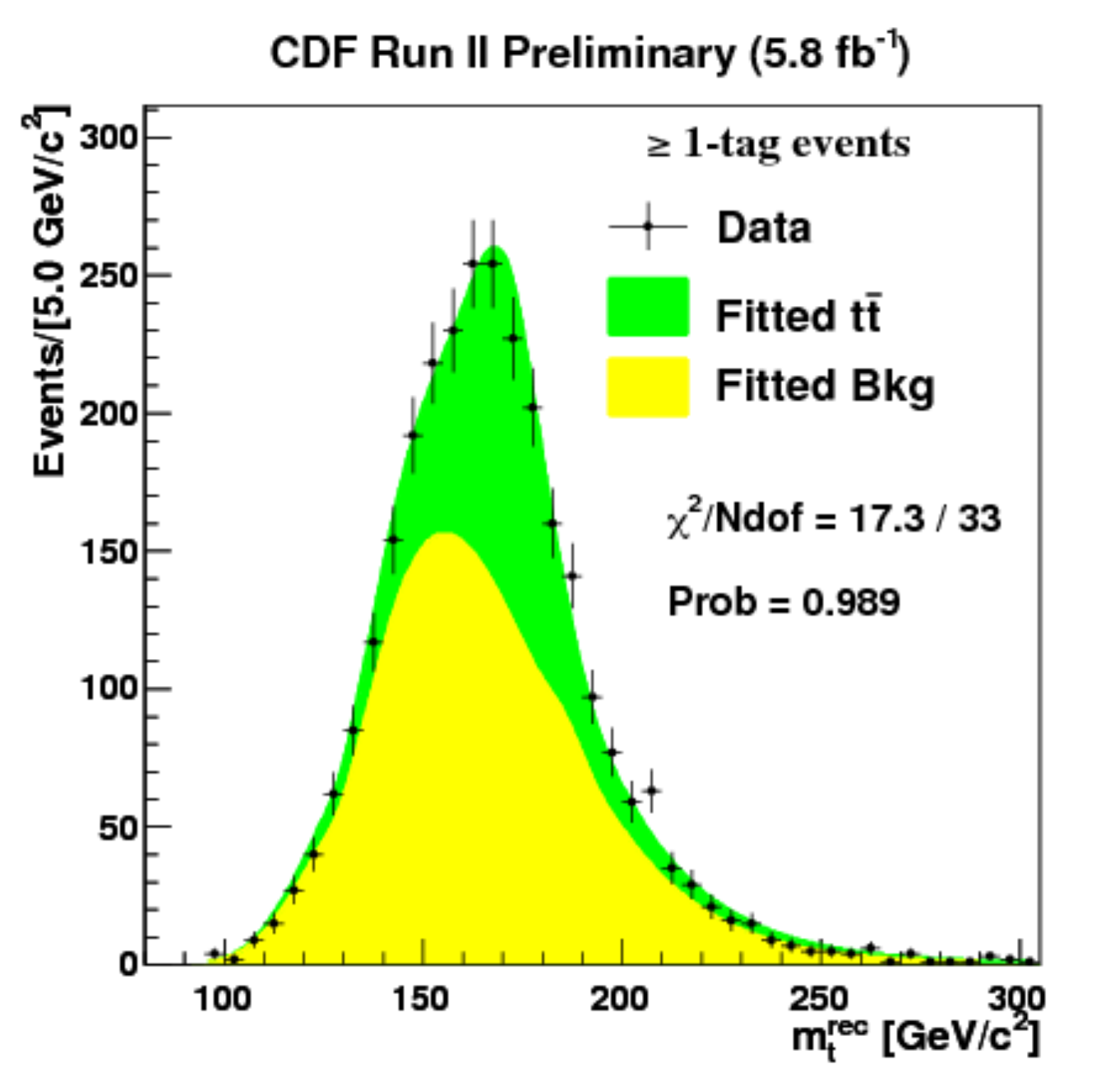}}
\caption{ Reconstructed top-quark mass for CDF all-jets data (5.8 fb$^{-1}$) with $\ge 1$ $b$-jets.}
\label{cdf-allj}
\end{figure}

The selection of the all-jets channel is quite difficult because of the huge QCD background expected and the difficulties in modeling the background. CDF applies  a neural-network based kinematical selection and a data-based modeling of the background. The measurement\cite{cdf-allj-ref}, obtained using a {\em template} method plus {\em in-situ} JES calibration, gives $M_t=172.5\pm 1.4({\rm stat.})\pm 1.0({\rm JES})\pm 1.1({\rm syst.})$ GeV, with a total uncertainty of 2.0 GeV. The systematic uncertainty is dominated by contributions due to the background modeling (0.6 GeV), the generator modeling (0.5 GeV) and the residual JES (0.4 GeV).  The expected and observed distribution of the reconstructed top-quark mass for events with $\ge 1$ $b$-jets are shown in Fig.~\ref{cdf-allj}.

\subsection{Tevatron average}
A big effort went into computing the average of all top-quark mass measurements in different channels. It is necessary in fact to evaluate all possible correlations among the various systematic uncertainties. Crucial is then a precise and common categorization of the contributions, allowing typically for a 0 or 50 or 100\% correlation between uncertainties for different channels and different experiments.
\par
\begin{figure}[hb]
\centerline{\includegraphics[width=8.8cm]{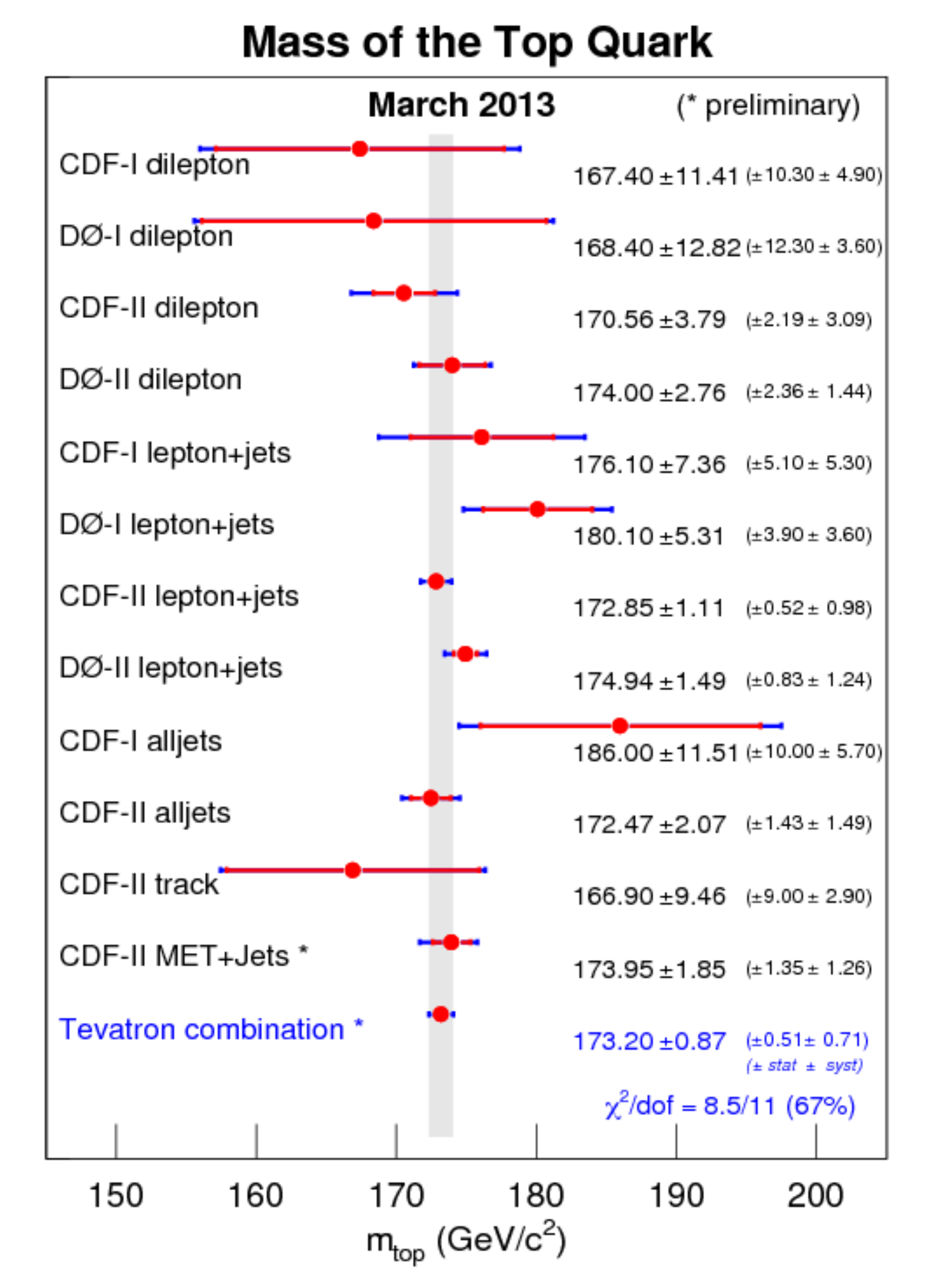}}
\caption{ Tevatron measurements of the top-quark mass in several channels and March 2013 average.}
\label{tev-ave}
\end{figure}

The average is computed with the Best Linear Unbiased Estimator (BLUE\cite{blue-ref1, blue-ref2}) assuming symmetric Gaussian uncertainties. The BLUE calculation performs a linear combination of the input measurements, optimizing the coefficients to be used by minimizing the uncertainty of the combined result. The algorithm considers both statistical and systematic uncertainties, and all uncertainties are assumed to be Gaussian-distributed.  The outcome of the average, at the time of this Conference, is\cite{tev-ave-ref}
$M_t^{TEV}=173.20\pm 0.51({\rm stat.})\pm 0.71({\rm syst.})$ GeV, with a total uncertainty of 0.87 GeV corresponding to 0.5\% of the mass itself. The various contributions to this average are shown in Fig.~\ref{tev-ave}.

\section{Top-Quark Mass Measurements at the LHC}
With the LHC turn-on additional events containing top quarks have been recorded, with integrated luminosities up to 5.0  fb$^{-1}$. Given the large increase in the size of the signal sample with respect to the Tevatron,  special care has been dedicated to the reduction of the systematic uncertainties.
\par
 We summarize in the following  the most recent (at the time of this Conference) measurements by ATLAS and CMS, in the usual $t\bar t$ channels.
\subsection{Lepton+jets channel}
The lepton+jets channel guarantees the most accurate measurements also at the LHC.
\par

\begin{figure}[hb]
\centerline{\includegraphics[width=6.8cm]{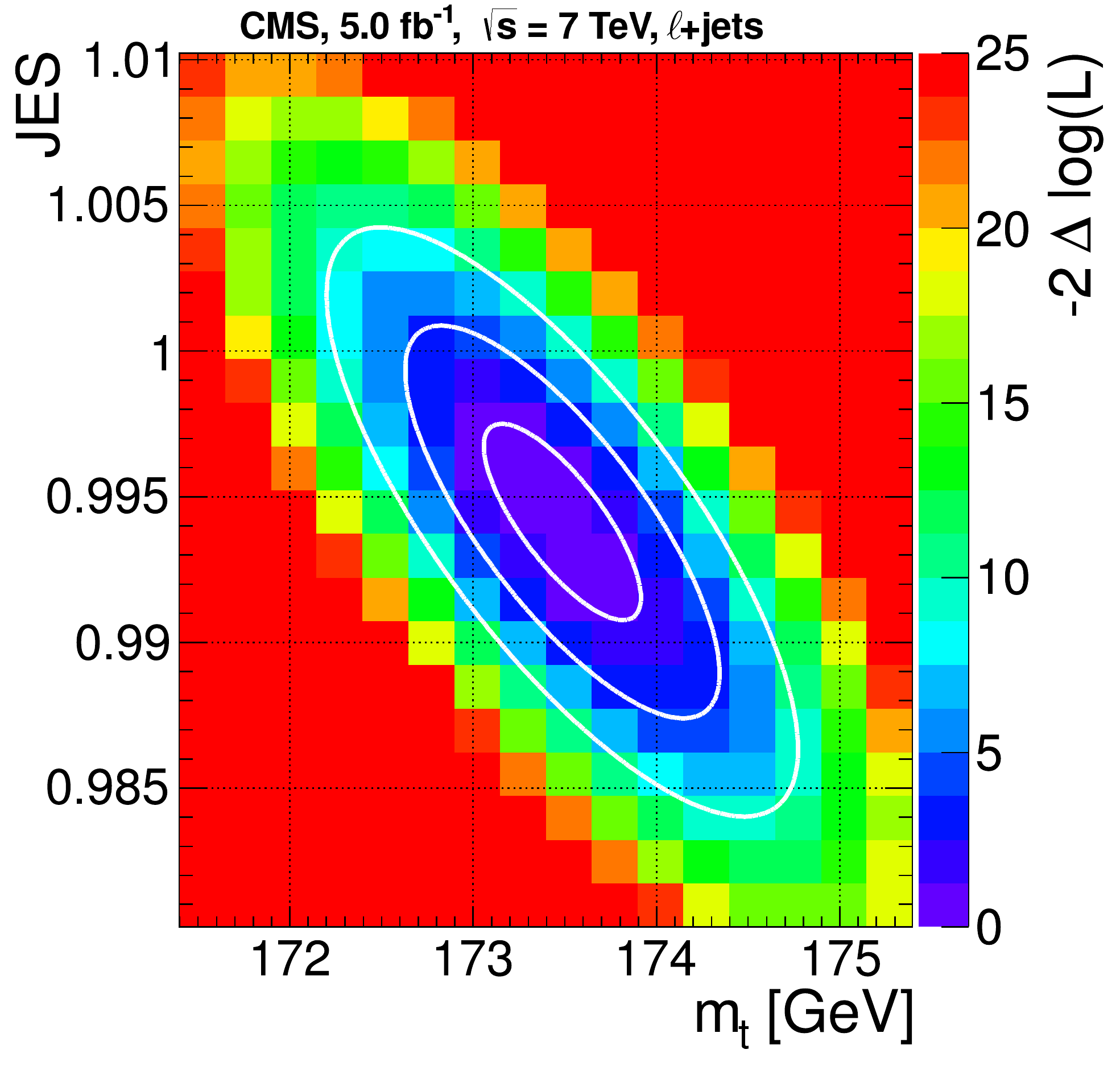}}
\caption{ 2D likelihood as a function of $M_t$ and JSF (named JES in the plot) for CMS lepton+jets data (5.0 fb$^{-1}$).}
\label{cms-lj}
\end{figure}

Applying the ideogram method with {\em in-situ} JES calibration, CMS measures\cite{cms-lj-ref} a top-quark mass $M_t=173.49\pm 0.43({\rm stat.+JES})\pm 0.98({\rm syst.})$ GeV, with a total uncertainty of 1.1 GeV. The systematic uncertainty is dominated by contributions due to the bJES (0.6 GeV), color-reconnection modeling (0.54 GeV) and the residual JES (0.28 GeV). The 2D likelihood as a function of the top-quark mass is shown in Fig.~\ref{cms-lj}.
\par
\begin{figure}[hb]
\begin{tabular}{cccc}
   \begin{minipage}{0.4\textwidth}
      \includegraphics[width=6.8cm]{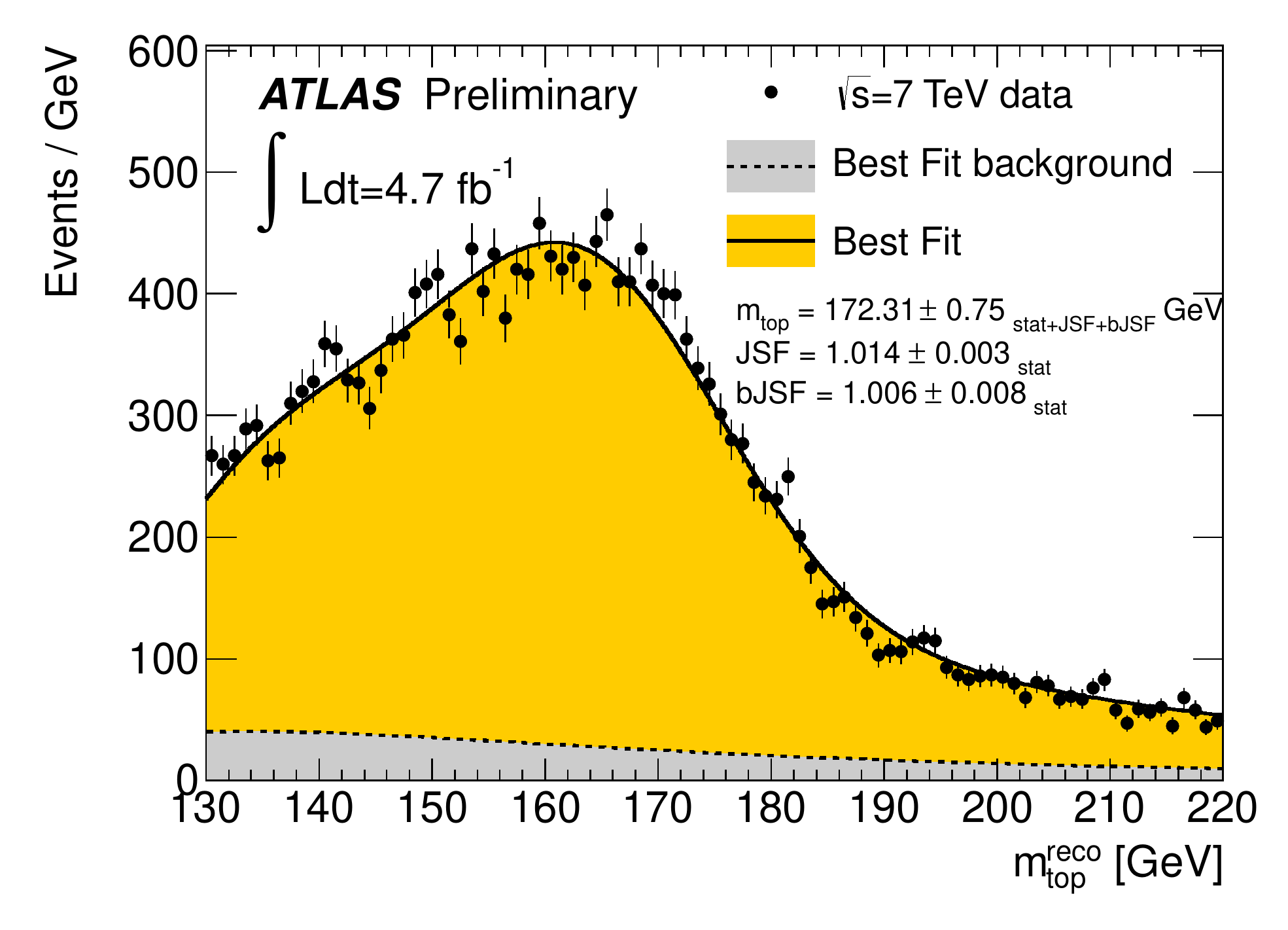}
   \end{minipage}
  & ~~~ & ~~~ &
   \begin{minipage}{0.4\textwidth} 
      \includegraphics[width=6.8cm]{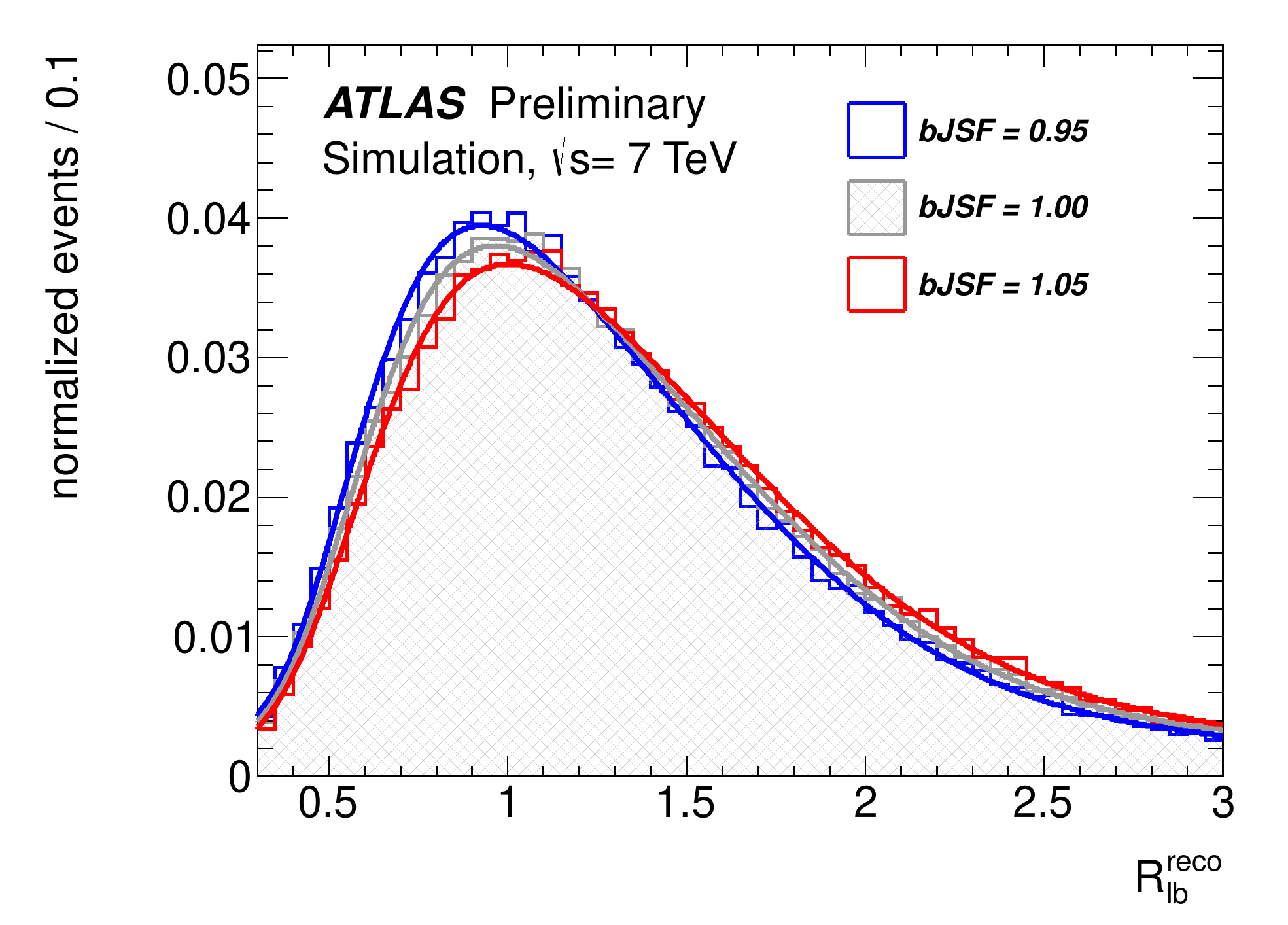}
   \end{minipage}
\end{tabular}
\caption{(Left) Reconstructed top-quark mass for ATLAS lepton+jets data (4.7 fb$^{-1}$) and best fit. (Right) Distribution for the bJSF for ATLAS simulations of lepton+jets data.}
\label{atlas-lj}
\end{figure}

A novelty introduced by ATLAS is the {\em in-situ} calibration also of the bJES by introducing a quantity, derived from the $p_T$ of untagged and tagged jets, which is sensitive to shifts in the bJSF value. The top-quark mass in this case is measured\cite{atlas-lj-ref} using a 3D template method and amounts to $M_t=172.31\pm 0.75({\rm stat.+JES+bJES})\pm 1.35({\rm syst.})$ GeV, with a total uncertainty of 1.5 GeV. The systematic uncertainty is dominated by contributions due to the $b$-jet tagging efficiency (0.81 GeV), the residual JES (0.79 GeV) uncertainty, and ISR/FSR effects (0.45 GeV). This new application of the bJES calibration was crucial to improve the total systematic uncertainty of the measurement from the previous value of 2.02 GeV to 1.35 GeV. The observed and expected distributions for the reconstructed top-quark mass are shown in Fig.~\ref{atlas-lj} (left), while in Fig.~\ref{atlas-lj} (right) is shown the bJSF distribution

\subsection{Dilepton channel}
\begin{figure}[hb]
\begin{tabular}{cccc}
   \begin{minipage}{0.4\textwidth}
      \includegraphics[width=6.8cm]{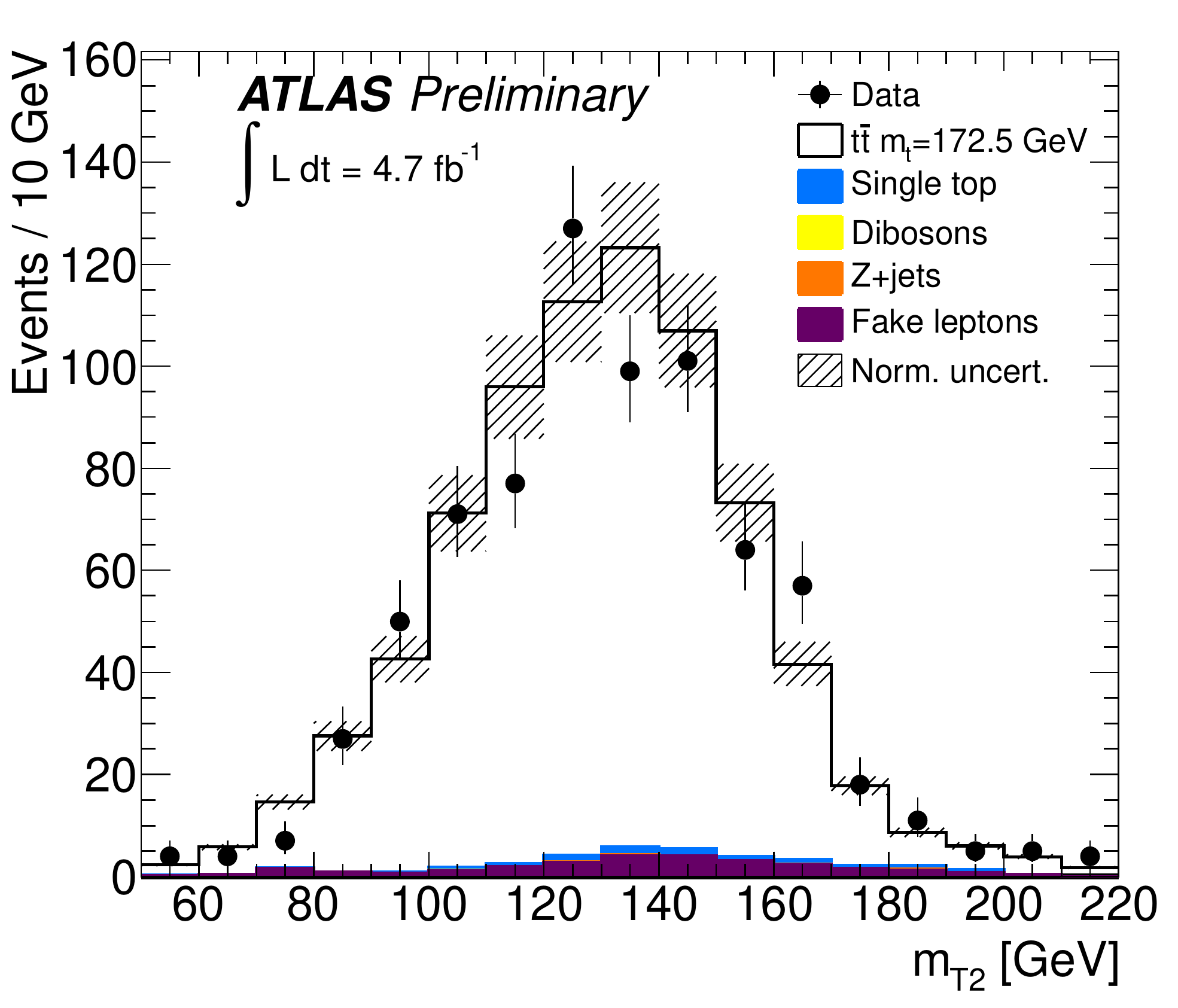}
   \end{minipage}
  & ~~~ & ~~~ &
   \begin{minipage}{0.4\textwidth} 
      \includegraphics[width=6.8cm]{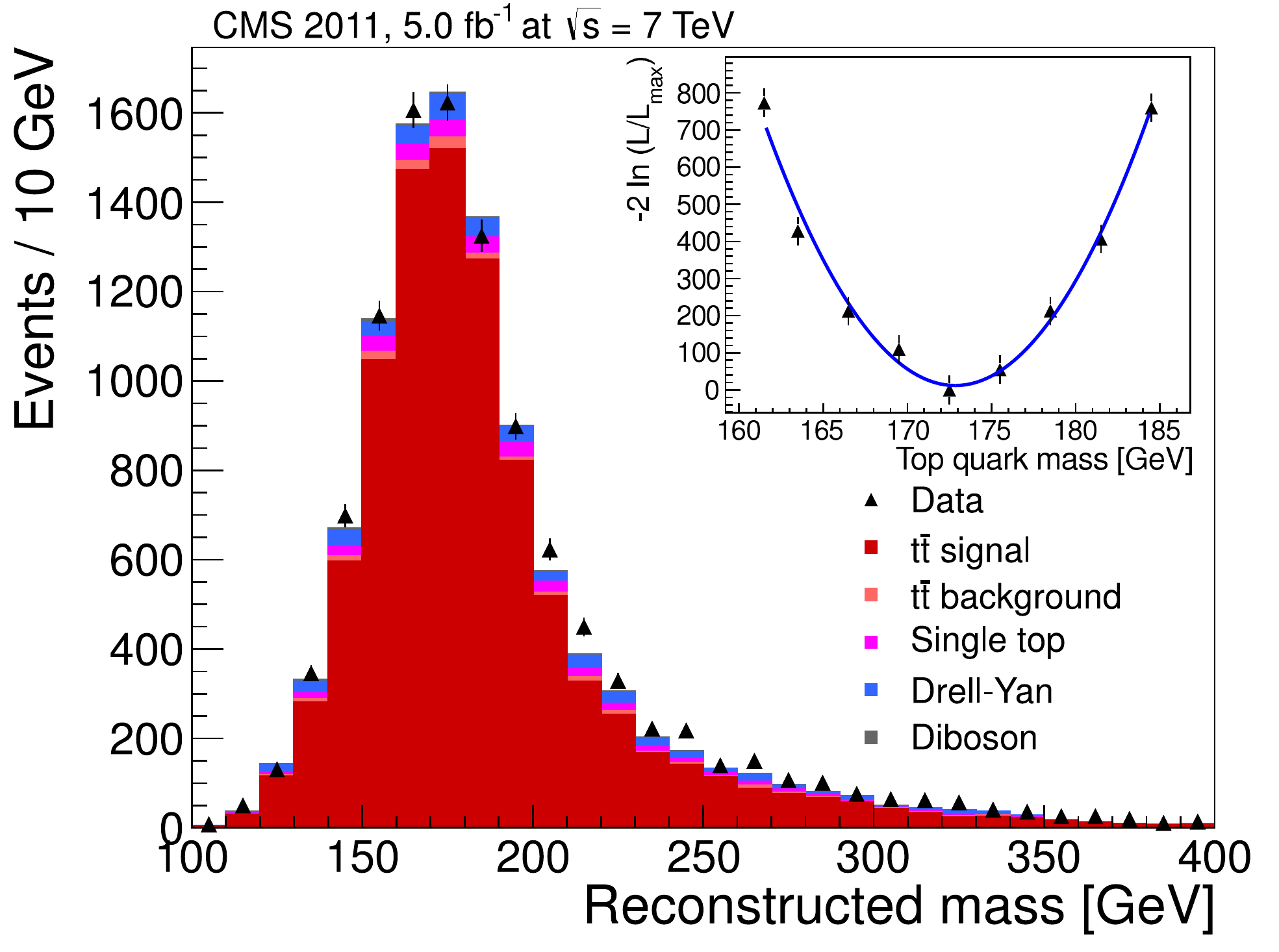}
   \end{minipage}
\end{tabular}
\caption{(Left) Distribution of the $m_{T2}$ variable for ATLAS dilepton data (4.7 fb$^{-1}$). (Right) Reconstructed top-quark mass for CMS dilepton data (5.0 fb$^{-1}$).}
\label{atlas-cms-dil}
\end{figure}

ATLAS measures $M_t$ in the dilepton channel by employing the $m_{T2}$ variable, also known as {\em stransverse mass}, which is usually considered in exotic searches for events with two undetected particles, and corresponds to the lower bound on the parent particle mass. The value measured\cite{atlas-dil-ref} is $M_t=175.2\pm 1.6({\rm stat.})^{+3.1}_{-2.8}({\rm syst.})$ GeV, with a total uncertainty of 3.5 GeV.  The major contributions to the systematic uncertainty come from the generator modeling (1.3 GeV), the JES (1.5 GeV) and the bJES (1.4 GeV) uncertainties. The observed and expected distributions for the $m_{T2}$ variable are shown in Fig.~\ref{atlas-cms-dil} (left).
\par
In the case of CMS the missing neutrinos are handled with an analytical matrix-weighting technique\cite{cms-dil-ref} which gives $M_t=172.5\pm 0.4({\rm stat.})\pm 1.5({\rm syst.})$ GeV, with a total uncertainty of 1.6 GeV.  The major contributions to the systematic uncertainty come from the JES (1.0 GeV) and the bJES (0.6 GeV) uncertainties. The observed and expected distributions for the reconstructed top-quark mass are shown in Fig.~\ref{atlas-cms-dil} (right).

\subsection{All-jets channel}
\begin{figure}[hb]
\begin{tabular}{cccc}
   \begin{minipage}{0.4\textwidth}
      \includegraphics[width=6.8cm]{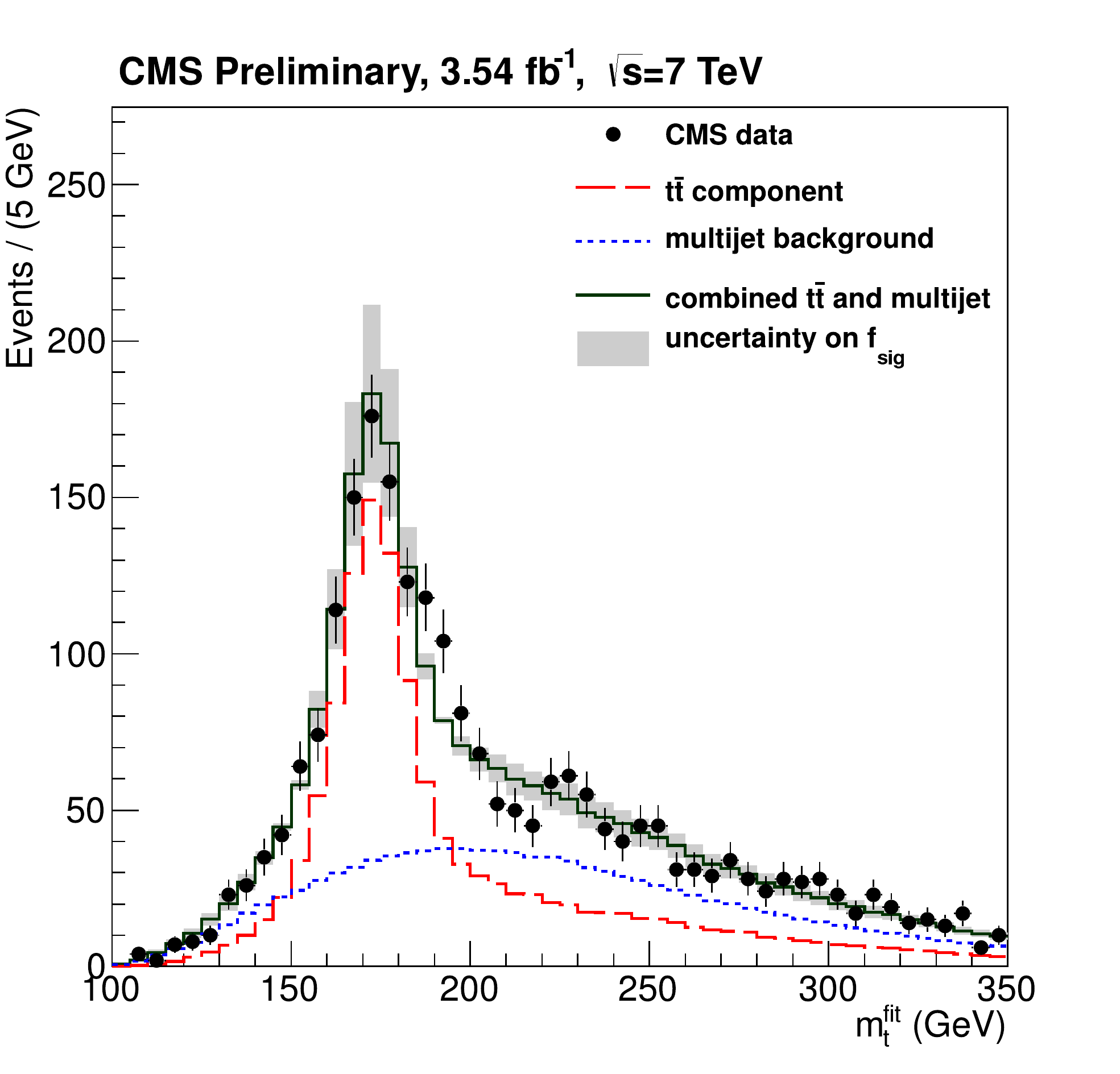}
   \end{minipage}
  & ~~~ & ~~~ &
   \begin{minipage}{0.4\textwidth} 
      \includegraphics[width=6.8cm]{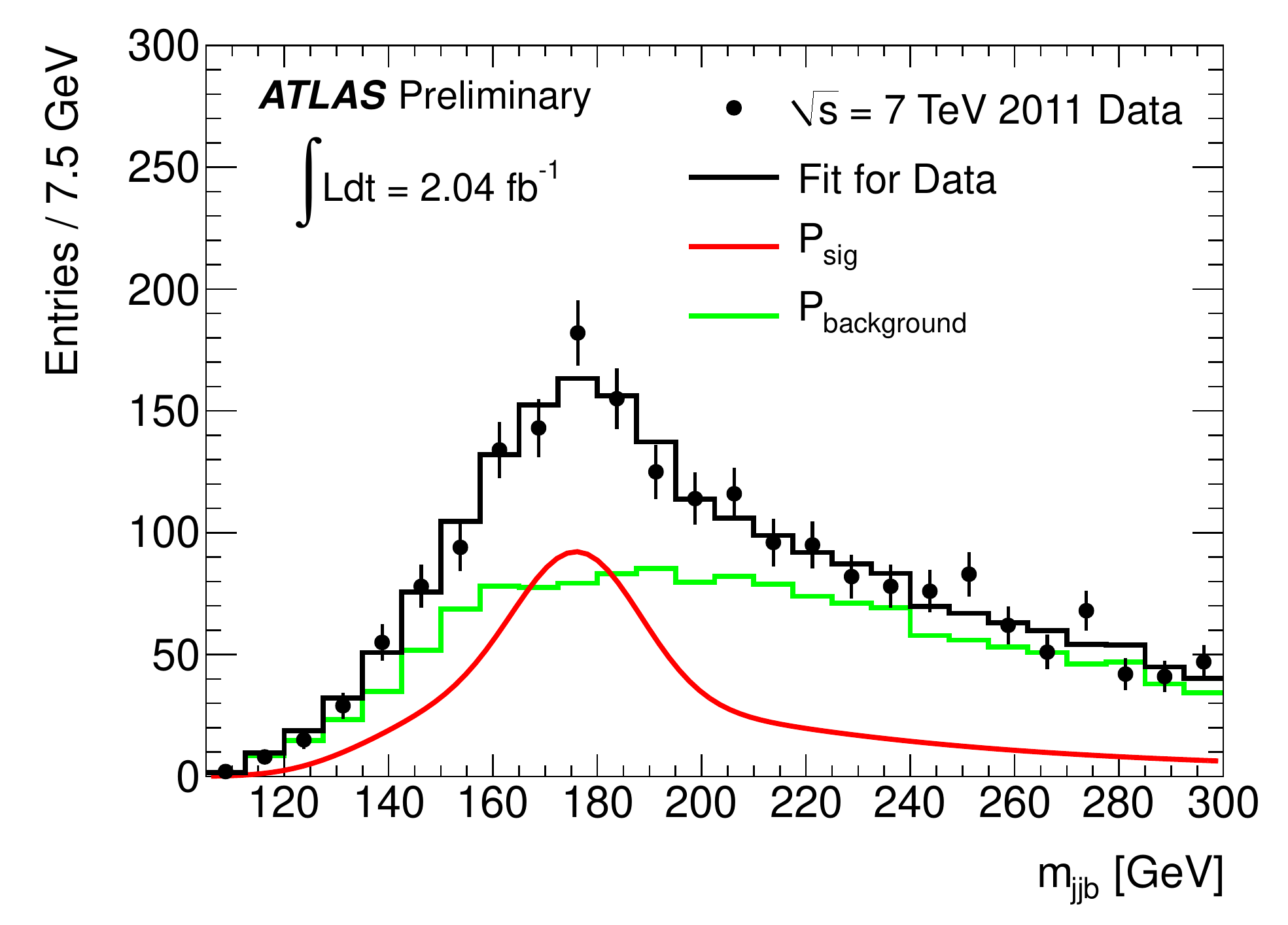}
   \end{minipage}
\end{tabular}
\caption{(Left) Distribution of the reconstructed top-quark mass for CMS all-jets data (3.5 fb$^{-1}$). (Right) Reconstructed top-quark mass for ATLAS all-jets data (2.04 fb$^{-1}$).}
\label{atlas-cms-allj}
\end{figure}

The challenging all-jets channel is pursued also at the LHC. With the ideogram method, but with no {\em in-situ} JES calibration, CMS measures\cite{cms-allj-ref} $M_t=173.49\pm 0.69({\rm stat.})\pm 1.25({\rm syst.})$ GeV, with a total uncertainty of 1.5 GeV.  The systematic uncertainty is dominated by contributions from the JES (0.97 GeV) and the bJES (0.49 GeV) uncertainties, and from the modeling of the underlying event (0.32 GeV). The observed and expected distributions for the reconstructed top-quark mass are shown in  Fig.~\ref{atlas-cms-allj} (left) .
\par
ATLAS applies a  template method and measures\cite{atlas-allj-ref} $M_t=174.9\pm 2.1({\rm stat.})\pm 3.8({\rm syst.})$ GeV, with a total uncertainty of 4.3 GeV. The systematic uncertainty is dominated by contributions from the JES (2.1 GeV) and the bJES (1.4 GeV), and from the modeling of the background (1.9 GeV). Distributions for the  observed and expected reconstructed top-quark mass are shown in  Fig.~\ref{atlas-cms-allj} (right).

\subsection{LHC average}
\begin{figure}[hb]
\centerline{\includegraphics[width=9.8cm]{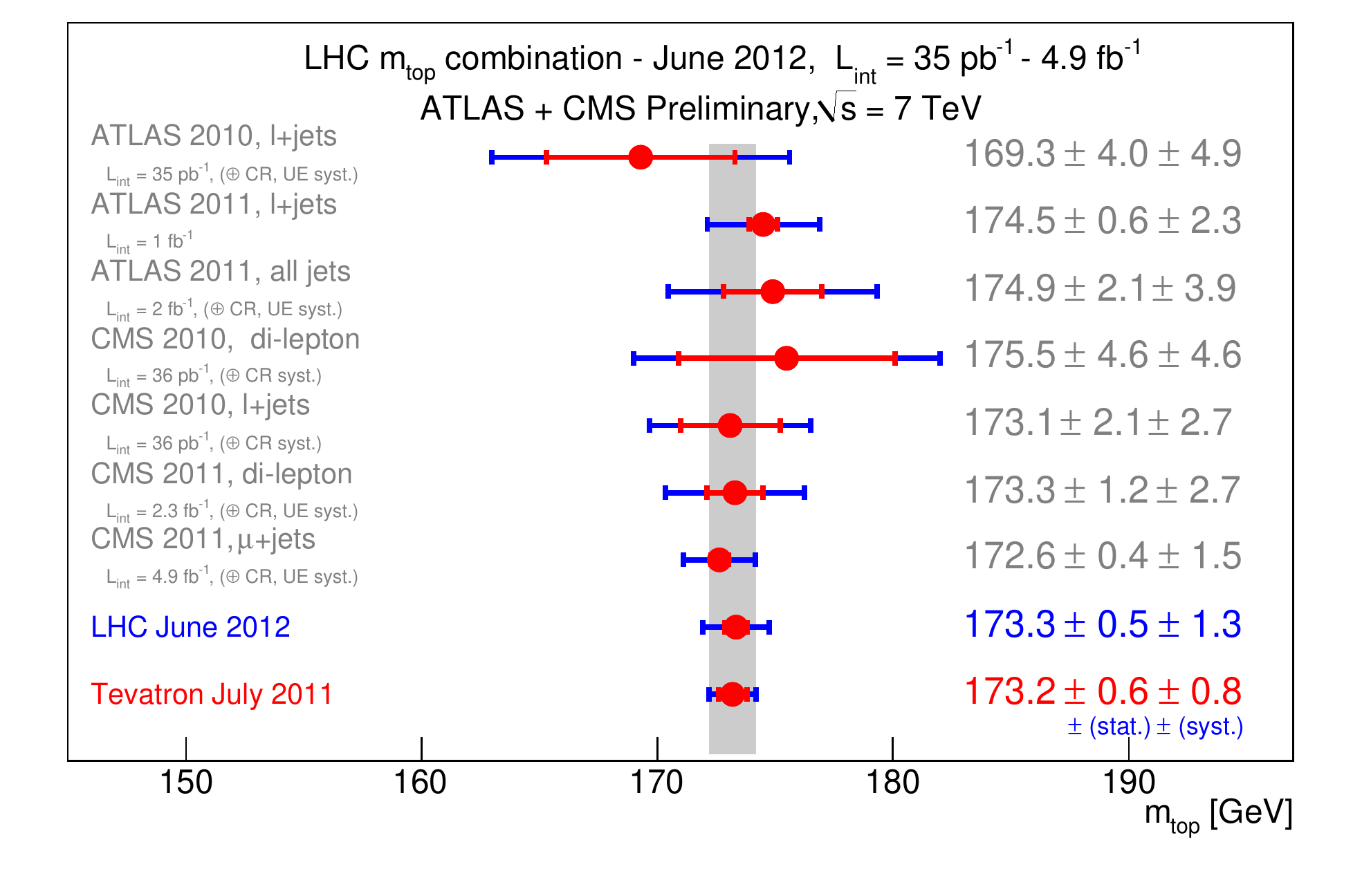}}
\caption{LHC measurements of the top-quark mass in several channels and June 2012 average.}
\label{lhc-ave}
\end{figure}

Also the LHC average of the available measurements is performed with the BLUE method, but a common and agreed definition of all the systematic uncertainties is in progress. The average available at the time of this Conference refers to 2012 results, although individual measurements discussed above are more precise than what used for this average. The 2012 LHC average\cite{lhc-ave-ref} is $M_t^{LHC}=173.3\pm 0.5({\rm stat.})\pm 1.3({\rm syst.})$ GeV, with a total uncertainty of 1.4 GeV corresponding to 0.8\% of the mass itself. The various contributions to the average are shown in Fig.~\ref{lhc-ave}. 

\section{Latest world average}
\begin{figure}[hb]
\centerline{\includegraphics[width=9.8cm]{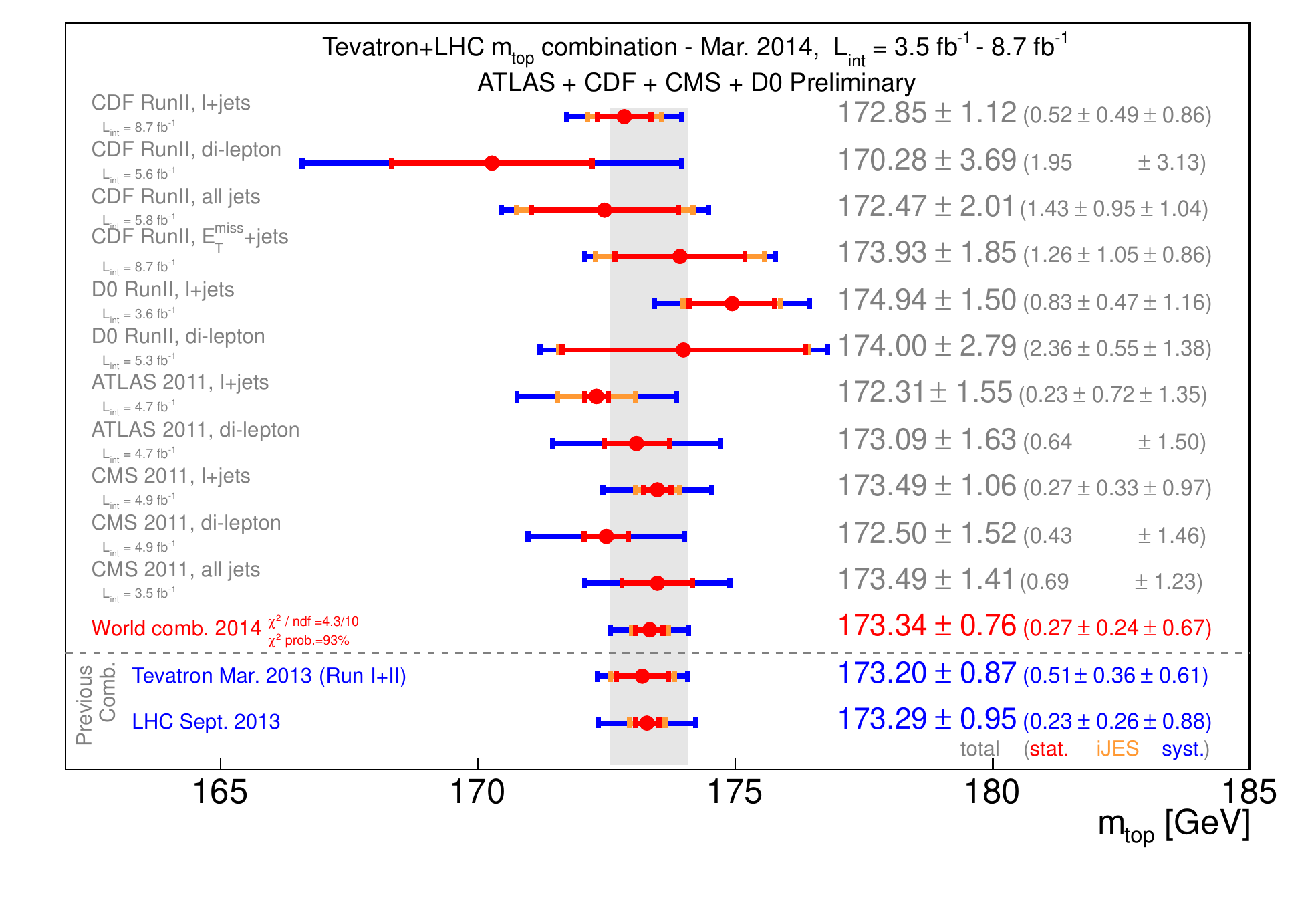}}
\caption{Tevatron+LHC March 2014 combination for the top-quark mass.}
\label{world-ave}
\end{figure}

A few months after this Conference, a combination of the latest measurements from the four experiments has been approved (March 2014) with a resulting value $M_t^{TEV+LHC}=173.34\pm 0.27({\rm stat.})\pm 0.71({\rm syst.})$ GeV. The total uncertainty of 0.76 GeV corresponds to a precision of only 0.44\%. We do not give any detail here but summarize the combination in Fig.~\ref{world-ave}. Full details can be found in Ref.~\refcite{latest-world-ave}.

\section{Top Quark versus Top Antiquark mass}
CPT invariance ensures that top quarks and antiquarks have the same mass. This principle can be proven experimentally by measuring the mass difference between $t$ and $\bar t$ quarks for instance using $t\bar t$ candidates in the lepton+jets channel.
\par
\begin{figure}[hb]
\begin{tabular}{cccc}
   \begin{minipage}{0.4\textwidth}
      \includegraphics[width=6.8cm]{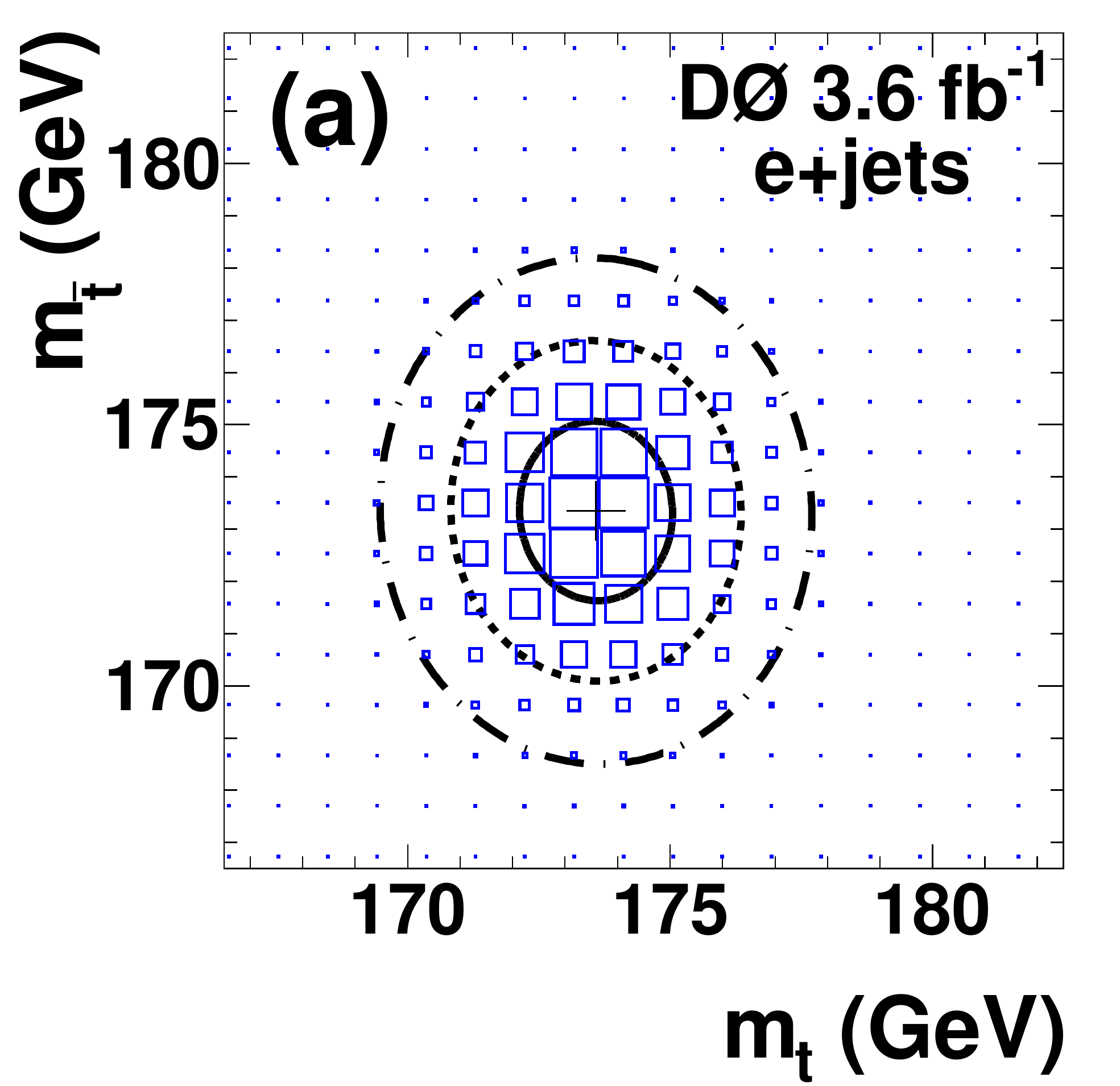}
   \end{minipage}
  & ~~~ & ~~~ &
   \begin{minipage}{0.4\textwidth} 
      \includegraphics[width=6.8cm]{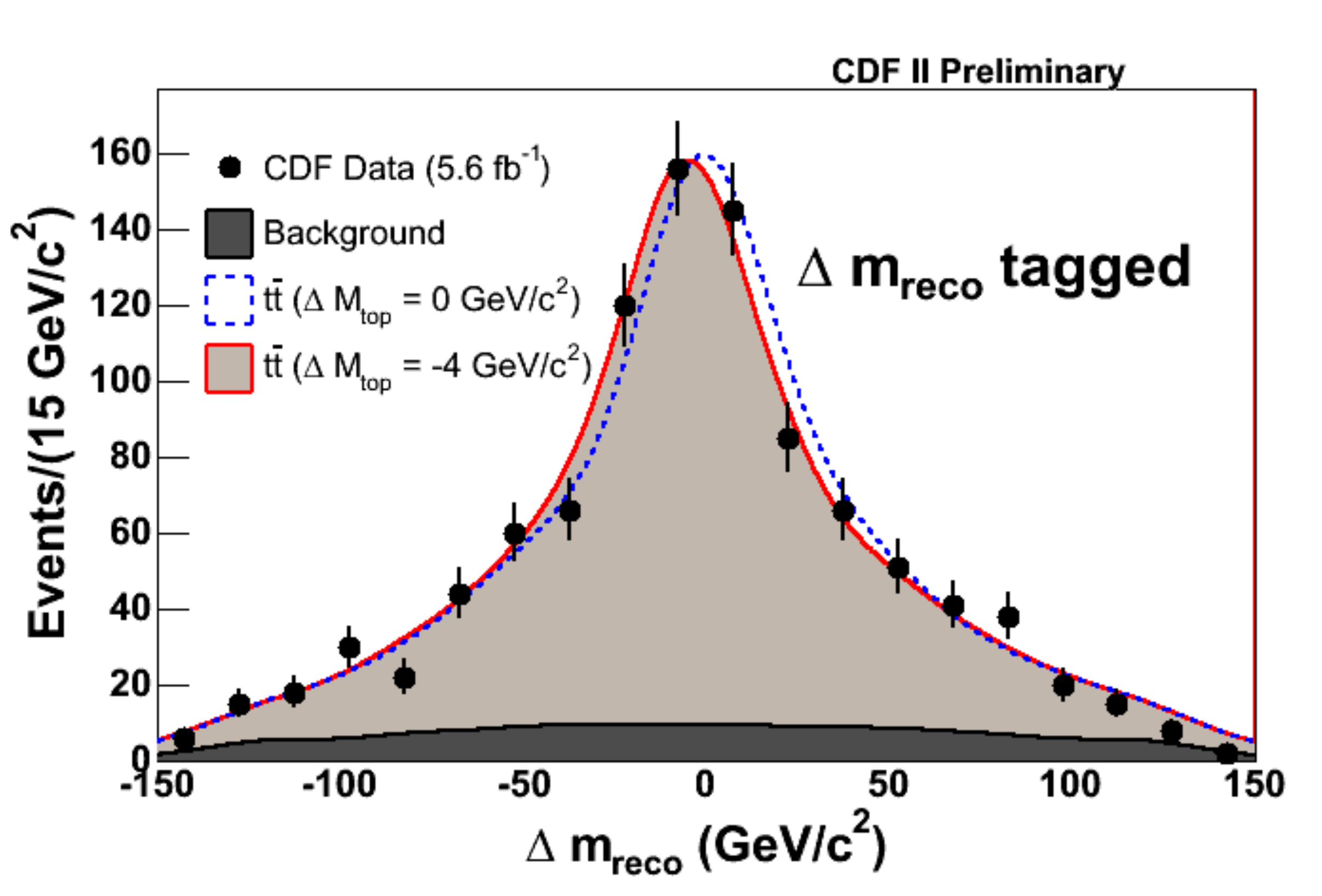}
   \end{minipage}
\end{tabular}
\caption{(Left) 2D likelihood densities as a function of $M_t$ and $M_{\bar t}$ for D0 lepton+jets data ($e$+jets, 3.6 fb$^{-1}$). (Right) Distribution of the mass difference $\Delta M_t$ for CDF lepton+jets data (8.7 fb$^{-1}$) with at least one $b$-jet.} 
\label{cdf-d0-diff}
\end{figure}

D0 uses a matrix element technique and the combination of $e$+jets and $\mu$+jets events gives a value for the mass difference, $\Delta M_t=M_t - M_{\bar t}$, which amounts to $0.8\pm 1.8({\rm stat.})\pm 0.5({\rm syst.})$, consistent with being 0. The 2D likelihood densities in terms of $M_t$ and $M_{\bar t}$ for $e$+jets events are shown in  Fig.~\ref{cdf-d0-diff} (left). A similar plot is obtained for $\mu$+jets events.
\par
CDF constructs templates for the mass difference and measures\cite{cdf-diff-ref}  $\Delta M_t=-1.95\pm 1.11 ({\rm stat.})\pm 0.59({\rm syst.})$ GeV, consistent with no difference. Observed and expected distributions for $\Delta M_t$ are shown in Fig.~\ref{cdf-d0-diff} (right).
\par
The test of CPT invariance is pursued also at the LHC where the experiments take advantage of the huge reduction in statistical uncertainty guaranteed by the larger $t\bar t$ yield.
\par
\begin{figure}[hb]
\begin{tabular}{cc}
\includegraphics[width=6.8cm]{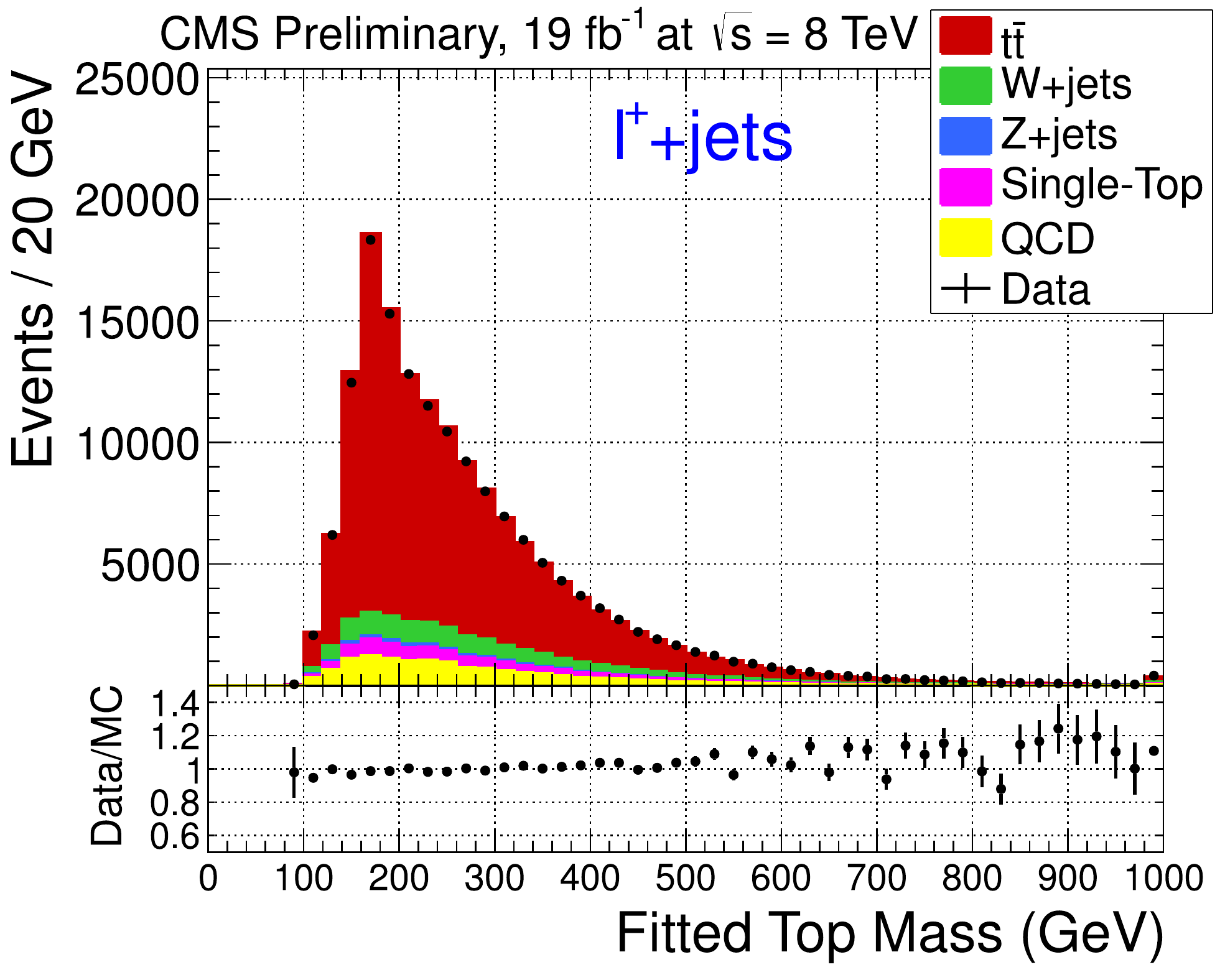}
\includegraphics[width=6.8cm]{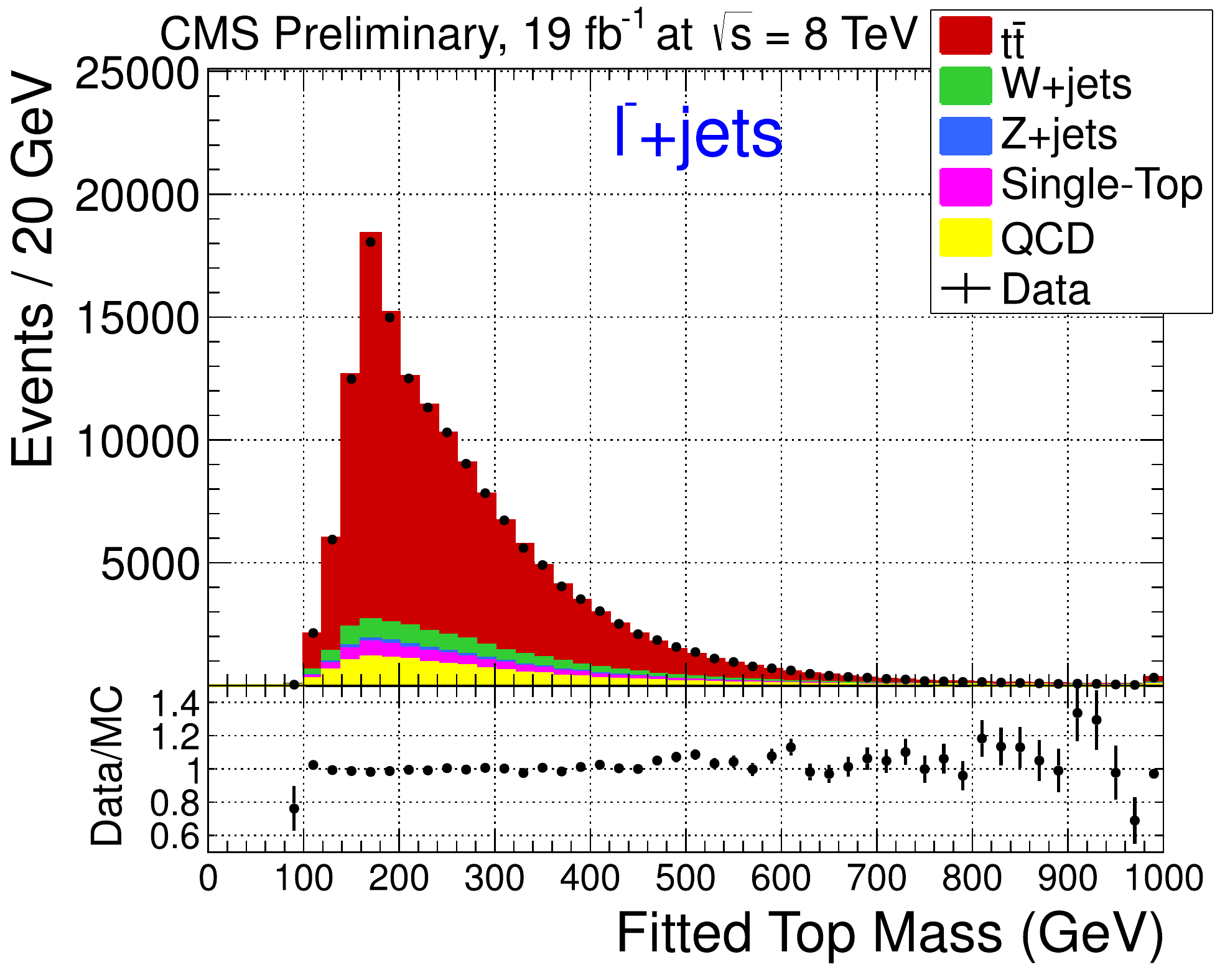}
\end{tabular}
\caption{Fitted top-quark masses for CMS lepton+jets data (19 fb$^{-1}$), for positive (left) or negative (right) charged leptons.}
\label{cms-diff}
\end{figure}

Applying an ideogram method CMS measures\cite{cms-diff-ref} $\Delta M_t=-272\pm 196 ({\rm stat.})\pm 122({\rm syst.})$ MeV, where the major contribution to the systematic uncertainty come from differences in $b$ vs $\bar b$ response (64 MeV) or uncertainties in the background composition (50 MeV). The value of the mass difference is again consistent with 0. Observed and expected distributions for top-quark masses associated to positive or negative leptons are shown in Fig.~\ref{cms-diff} (left).

\section{Theoretical Mass}
The level of precision reached in the direct measurement of the top-quark mass forces a reflection on the meaning of the measured quantity. In fact, the quantity whose measurements have been described above corresponds to the mass parameter, $M_t^{MC}$, used as input in the MC generation, which is typically evaluated at leading order or next-to-leading order.
\par
The extraction of $M_t$ is however affected by several perturbative and non-perturbative small ($< 1\%$) uncertainties associated to issues like:
\begin{itemize}
\item the modeling in the MC of the final state;
\item the reconstruction of the $t\bar t$ system;
\item the finite top-quark width, which affects the tails of kinematical distributions;
\item the presence of possible bound-state effects;
\item missing higher orders in the generation;
\item color-reconnection effects.
\end{itemize}
The increasing level of precision requires to relate $M_t$ to theory-based quantities like:
\begin{itemize}
\item the {\em pole mass}, $M_t^{pole}$, which is a universal quantity, but theoretically ambiguous by amounts of ${\cal O}(\Lambda_{QCD})$, due to soft gluon radiation (the so-called {\em infrared renormalon problem}\cite{renormalon-ref});
\item Lagrangian masses, which are theoretically unambiguous but not universal, like the $\overline{\rm MS}$ mass, which is defined only in perturbation theory.
\end{itemize}
For more details see for instance Refs.~\refcite{hoang-ref,ahrens-ref} for a discussion in the context of hadron colliders.
\par
These theory-oriented masses can be derived from a comparison of the measured production cross section $\sigma_{t\bar t}$ to theoretical predictions of  $\sigma_{t\bar t}$ vs $M_t$. Of course in the predictions one has to make assumptions on what the MC parameter $M_t^{MC}$ is equal to, because that value affects for instance  the estimation of the selection efficiencies.
\par
\begin{figure}[hb]
\begin{tabular}{cc}
\includegraphics[width=6.8cm]{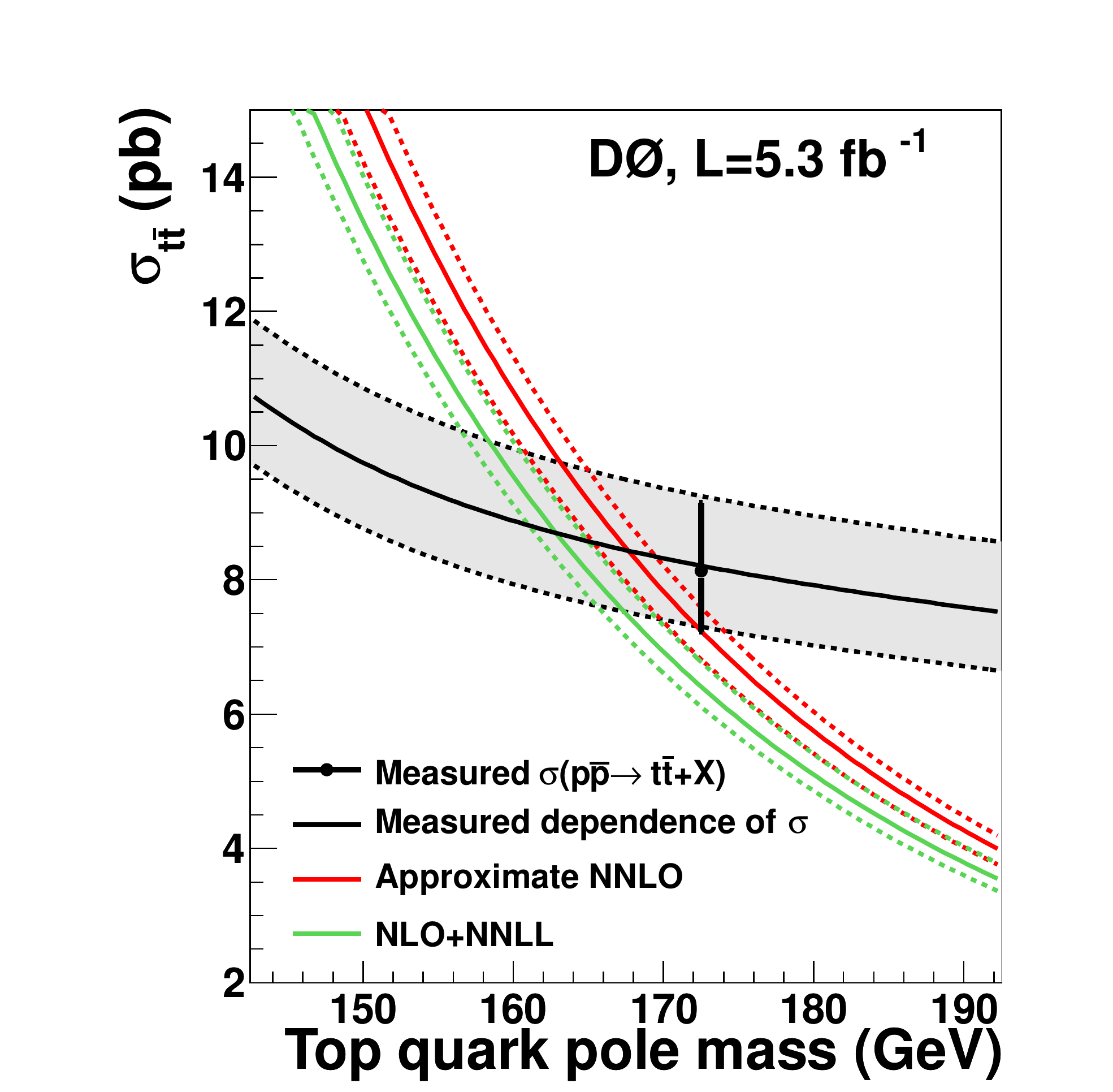}
\includegraphics[width=6.8cm]{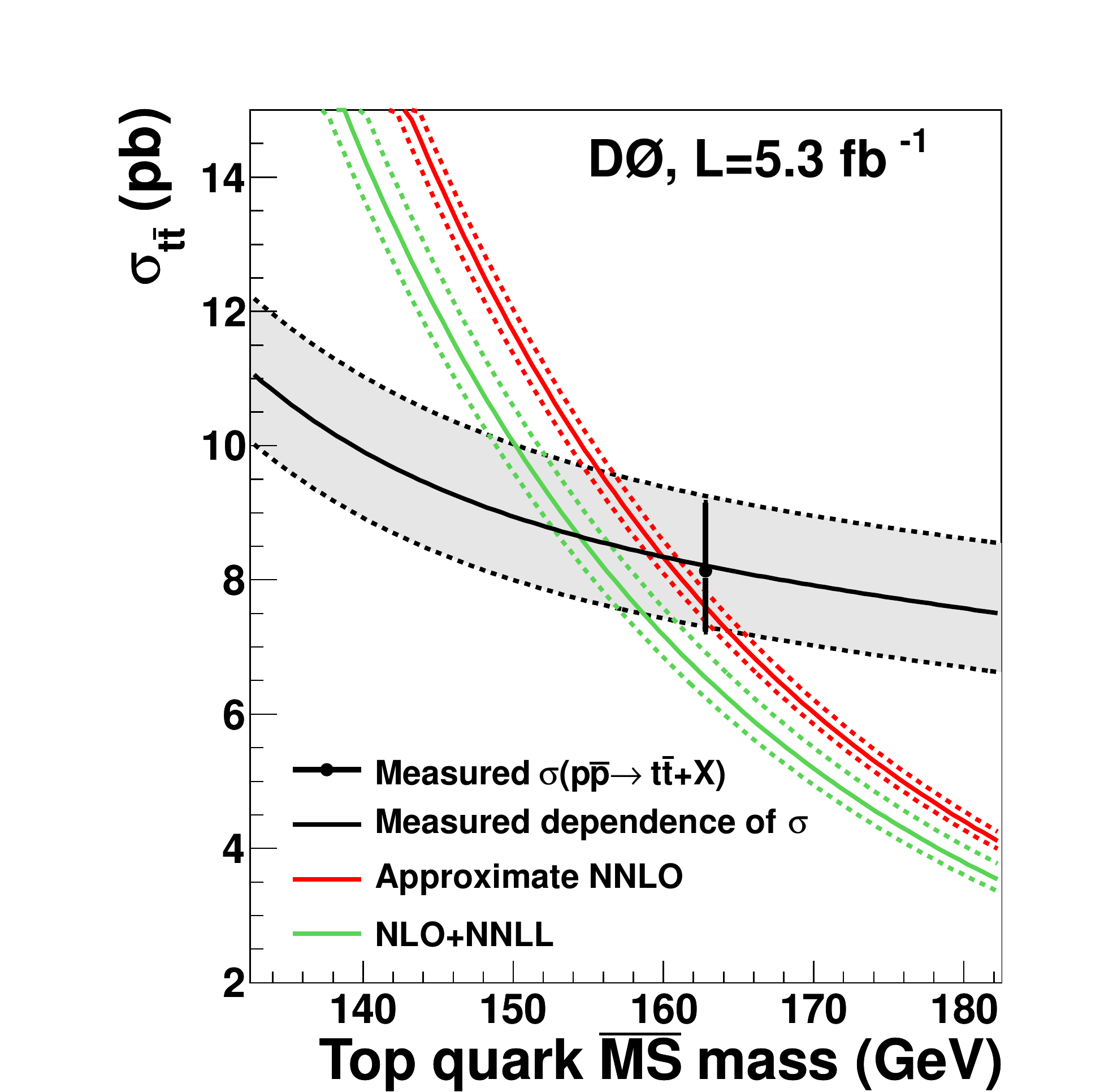}
\end{tabular}
\caption{Production cross section vs top-quark pole mass (left) or $\overline{\rm MS}$ mass (right) for D0 lepton+jets data (5.3 fb$^{-1}$). Theoretical predictions at various orders are shown.}
\label{d0-pole-ms}
\end{figure}

From a $\sigma_{t\bar t}$ measurement in the lepton+jets channel, and comparing it with theoretical predictions at different orders, D0 measured\cite{d0-polemass-ref} the pole mass to be approximately $165-167$ GeV (see Fig.~\ref{d0-pole-ms} (left)) assuming $M_t^{MC}=M_t^{pole}$ and using different theoretical calculations. The value of $M_t^{pole}$ is  consistent at $2\sigma$ level with the current world average\cite{world-ave-ref} $173.5\pm 1.0$ GeV for the direct measurement of the top-quark mass. As for the $\overline{\rm MS}$ values, they amount to about $155-160$ GeV (see Fig.~\ref{d0-pole-ms} (right)) which differ more than $2\sigma$ from the world average.
\par
\begin{figure}[hb]
\begin{tabular}{cccc}
   \begin{minipage}{0.4\textwidth}
      \includegraphics[width=6.8cm]{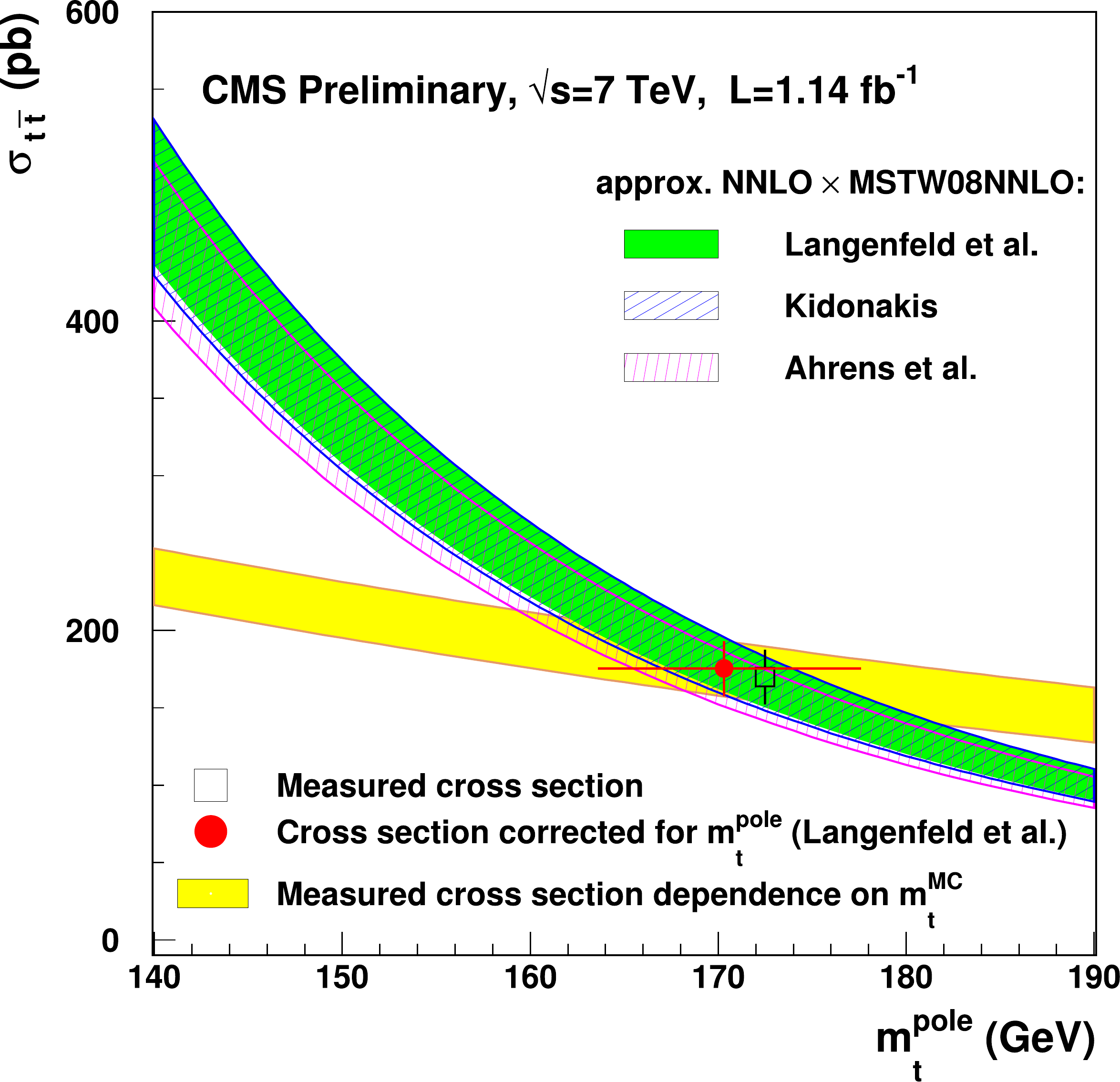}
   \end{minipage}
  & ~~~ & ~~~ &
   \begin{minipage}{0.4\textwidth} 
      \includegraphics[width=6.8cm]{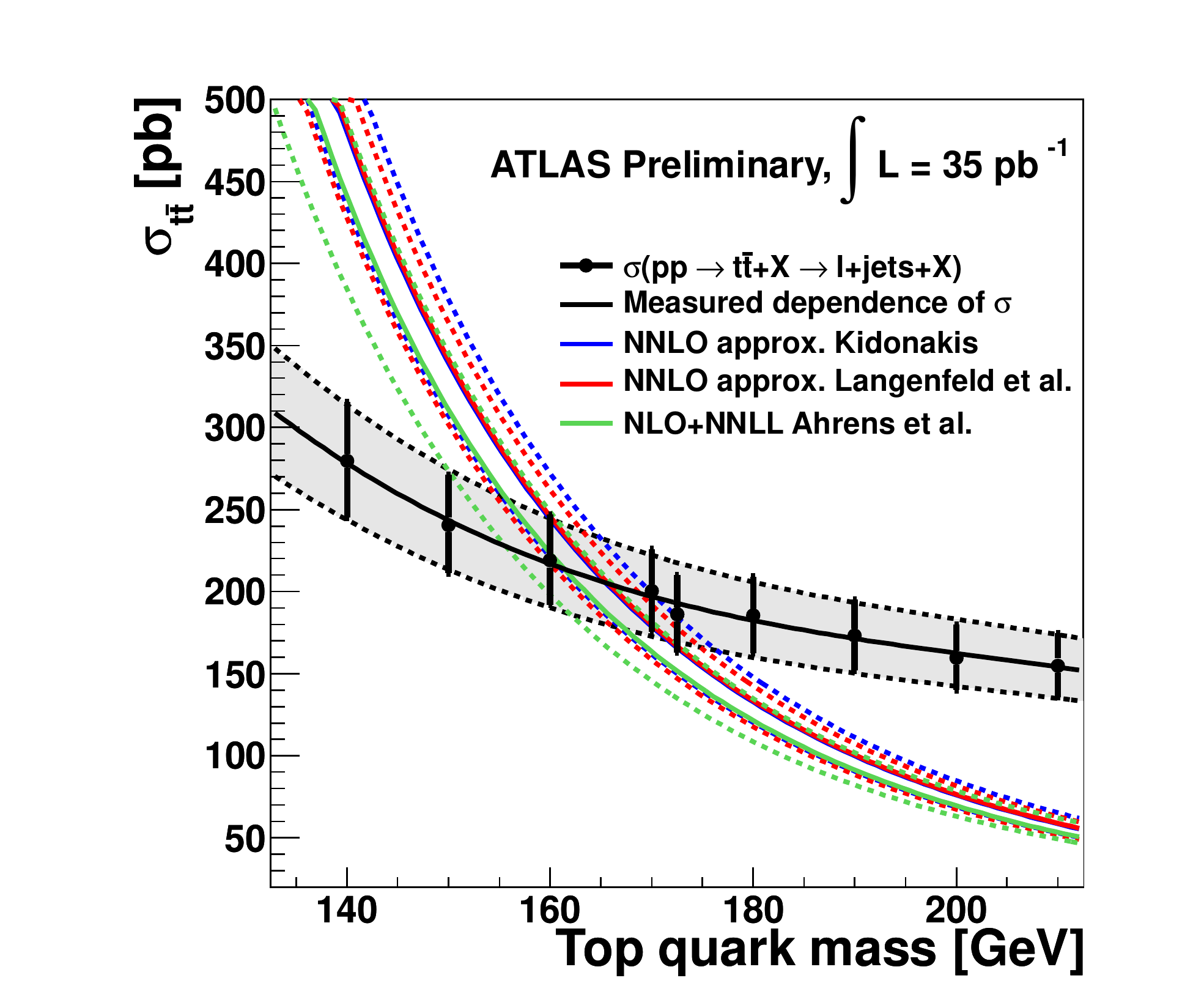}
   \end{minipage}
\end{tabular}
\caption{Production cross section vs top-quark pole mass for CMS lepton+jets data (1.1 fb$^{-1}$, left) and for ATLAS lepton+jets data (35 pb$^{-1}$, right). 
Theoretical predictions at various orders are shown\cite{lagen-xsec,kido-xsec,ahrens-xsec}.}
\label{lhc-pole-mass}
\end{figure}

Pole and $\overline{\rm MS}$ masses are measured also at LHC, in the lepton+jets channel. CMS measures\cite{cms-polemass-ref} a pole mass of about 170 GeV (see Fig.~\ref{lhc-pole-mass} (left)) 
and a  $\overline{\rm MS}$ mass of about 160 GeV, with an uncertainty of approximately 6 GeV.  ATLAS measures\cite{atlas-polemass-ref} a pole mass of about 166 GeV, with an uncertainty of approximately 8 GeV (see Fig.~\ref{lhc-pole-mass} (right)). Again, the level of sensitivity does not allow at present to discriminate between different theoretical calculations.
\section{Conclusions}
Since the discovery of the top quark, the measurement of its mass has been pursued in a variety of channels and with different techniques.
The level of precision reached in the measurement of the top-quark mass is impressive, $<0.5\%$, thanks to 18 years of continuous accumulation of data and improvements in the methodology. An even better precision is expected from ongoing and future measurements conducted at the LHC. New  measurements at increasing precision will  help to explore fundamental issues like:
\begin{itemlist}
\item cosmological models for inflation;
\item vacuum stability of the SM;
\item  physics beyond the SM.
\end{itemlist}
To achieve these goals it will be important to reduce the systematic uncertainties, mainly those related to signal modeling, by improving the tuning of the parameters in the MC generators and their agreement with the data.

\section*{Acknowledgments}
The measurements described here are the results of the work of several people in the 4 Collaborations, and  a special thank goes the top group convenors: T. Carli, M. J. Costa (ATLAS), D. Toback, J. Wilson (CDF), A. Meyer, M. Mulders, R. Tenchini (CMS), A. Jung, V. Sharyy (D0) for their help in putting all these results together.


\begin{thebibliography}{0}    

\bibitem{cdf-disc} CDF Collab. (F. Abe {\it et al}.), Phys. Rev. Lett. {\bf 74}, 2626 (1995), doi:10.1103/PhysRevLett.74.2626, arXiv:hep-ex/9503002.

\bibitem{d0-disc} D0 Collab. (S. Abachi {\it et al}.), Phys. Rev. Lett. {\bf 74}, 2632 (1995),  doi:10.1103/PhysRevLett.74.2632, arXiv:hep-ex/9503003.

\bibitem{top-xsec} M. Czakon, P. Fiedler and A. Mitov, arXiv:1303.6254 [hep-ph].

\bibitem{world-ave-ref} Particle Data Group, Phys. Rev. D {\bf 86}, 010001 (2012), doi:10.1103/PhysRevD.86.010001.


\bibitem{tev-review} A. B. Galtieri, F. Margaroli and I. Volobouev, Rept. Progr. Phys. {\bf 75} (2012) 056201, arXiv:1109.2163 [hep-ex].

\bibitem{lhc-review} F.-P. Schilling, 
 Int. J. Mod. Phys. A {\bf 27} (2012) 1230016 and arXiv:1206.4484 [hep-ex].

\bibitem{atlas-higgs-ref} ATLAS Collab. (G. Aad {\it et al}.), Phys. Lett. B {\bf 716}, 1 (2012), doi:10.1016/j.physletb.2012.08.020,  arXiv:1207.7214 [hep-ex].

\bibitem{cms-higgs-ref}  CMS Collab. (S. Chatrchyan {\it et al}.), Phys. Lett. B {\bf 716}, 30 (2012), doi:10.1016/j.physletb.2012.08.021,  arXiv:1207.7235 [hep-ex].

\bibitem{sm-global-fit} M. Baak, M. Goebel, J. Haller, A. Hoecker, D. Kennedy, R. Kogler, K. Moenig, M. Schott and J. Stelzer, Eur. Phys. J. C {\bf 72}, 2205 (2012), doi:10.1140/epjc/s10052-012-2205-9, arXiv:1209.2716 [hep-ph].

\bibitem{topcolor-ref1} C. T. Hill, Phys. Lett. B {\bf 266}, 419 (1991), doi: 10.1016/0370-2693(91)91061-Y.

\bibitem{topcolor-ref2} C. T. Hill, Phys. Lett. B {\bf 345}, 483 (1995), doi:10.1016/0370-2693(94)01660-5,  arXiv:hep-ph/9411426.

\bibitem{ewk-break-ref} W. A. Bardeen, C. T. Hill and M. Lindner, Phys. Rev. D {\bf 41}, 1647 (1990), doi:10.1103/PhysRevD.41.1647.


\bibitem{vacuum-stability} G. Degrassi, S. Di Vita, J. Elias-Mir\'o, J. R. Espinosa, G. F. Giudice, G. Isidori and A. Strumia, arXiv:1205.6497 [hep-ph].

\bibitem{ideogram-ref} DELPHI Collab. (P. Abreu {\it et al}.), Eur. Phys. J. C {\bf 2}, 581 (1998), doi:10.1007/s100520050163.

\bibitem{matrix-ref}  D0 Collab. (V. M. Abazov {\it et al}.), Nature {\bf 429}, 638 (2004),  doi:10.1038/nature02589, arXiv:hep-ex/0406031.

\bibitem{cdf-lj-ref} CDF Collab. (T. Aaltonen {\it et al}.), Phys. Rev. Lett. {\bf 109}, 152003 (2012), doi:10.1103/PhysRevLett.109.152003, arXiv:1207.6758 [hep-ex].

\bibitem{d0-lj-ref} D0 Collab. (V. M. Abazov {\it et al}.), Phys. Rev. D {\bf 84}, 032004 (2011),  doi:10.1103/PhysRevD.84.032004,  arXiv:1105.6287 [hep-ex].

\bibitem{d0-dil-ref} D0 Collab. (V. M. Abazov {\it et al}.), Phys. Rev. Lett. {107}, 082004 (2011), doi: 10.1103/PhysRevLett.107.082004,  arXiv:1105.0320 [hep-ex].

\bibitem{cdf-dil-ref} CDF Collab. (T. Aaltonen {\it et al}.), Phys. Rev. D {\bf 88}, 011101 (2013), doi:10.1103/PhysRevD.88.011101,  arXiv:1305.3339 [hep-ex].

\bibitem{cdf-allj-ref} CDF Collab. (T. Aaltonen {\it et al}.), Phys. Lett. B {\bf 714}, 24 (2012), doi:10.1016/j.physletb.2012.06.007,  arXiv:1112.4891 [hep-ex].

\bibitem{blue-ref1} L. Lyons, D. Gibaut and P. Clifford, Nucl. Instrum. Meth. A {\bf 270}, 110 (1998).

\bibitem{blue-ref2} A. Valassi,  Nucl. Instrum. Meth. A {\bf 500}, 391 (2003).

\bibitem{tev-ave-ref} Tevatron Electroweak Working Group, arXiv:1305.3929 [hep-ex].

\bibitem{cms-lj-ref} CMS Collab. (S. Chatrchyan {\it et al}.), JHEP {\bf 1212}, 105 (2012), doi:10.1007/JHEP12(2012)105,  arXiv:1209.2319 [hep-ex].

\bibitem{atlas-lj-ref} ATLAS Collab., Conference note ATLAS-CONF-2013-046 (2013), http://cdsweb.cern.ch/record/1547327.

\bibitem{atlas-dil-ref} ATLAS Collab., Conference note ATLAS-CONF-2012-082 (2013), http://cdsweb.cern.ch/record/1460394.


\bibitem{cms-dil-ref} CMS Collab. (S. Chatrchyan {\it et al}.), Eur. Phys. J. C {\bf 72}, 2202 (2012), doi:10.1140/epjc/s10052-012-2202-z,  arXiv:1209.2393 [hep-ex].


\bibitem{cms-allj-ref} CMS Collab., CMS Physics Analysis Summary CMS-PAS-11-017 (2012), http://cdsweb.cern.ch/record/1477721. 

\bibitem{atlas-allj-ref} ATLAS Collab., Conference note ATLAS-CONF-2012-030 (2012), http://cdsweb.cern.ch/record/1431895. 


\bibitem{lhc-ave-ref} ATLAS and CMS Collabs., Conference note ATLAS-CONF-2012-095 (2012) and CMS Physics Analysis Summary CMS-PAS-12-001 (2012), http://cdsweb.cern.ch/record/1460441. 

\bibitem{latest-world-ave} ATLAS, CDF, CMS and D0 Collabs., http://inspirehep.net/record/1286320, arXiv:1403.4427 [hep-ex].

\bibitem{d0-diff-ref} D0 Collab. (V. M. Abazov {\it et al}.), Phys. Rev. D {\bf 84}, 052005 (2011), doi:10.1103/PhysRevD.84.052005,  arXiv:1106.2063 [hep-ex]. 

\bibitem{cdf-diff-ref}  CDF Collab. (T. Aaltonen {\it et al}.), Phys. Rev. D {\bf 87}, 052013 (2013), doi:10.1103/PhysRevD.87.052013, arXiv:1210.6131 [hep-ex].

\bibitem{cms-diff-ref} CMS Collab., CMS Physics Analysis Summary CMS-PAS-12-031 (2013), http://cdsweb.cern.ch/record/1528156. 

\bibitem{renormalon-ref} A. H. Hoang, A. Jain, I. Scimemi and I. W. Stewart, Phys. Rev. Lett. {\bf 101}, 151602 (2008), doi:10.1103/PhysRevLett.101.151602 and  arXiv:0803.4214 [hep-ph].

\bibitem{hoang-ref} A. H. Hoang and I. W. Stewart, Nucl. Phys. Proc. Suppl {\bf 185}, 220 (2008), doi10.1016/j.nuclphysbps.2008.10.028,  arXiv:0808.0222 [hep-ph].

\bibitem{ahrens-ref} V. Ahrens, A. Ferroglia, M. Neubert, B. D. Pecjak and L. L. Yang, Phys. Lett. B {\bf 703}, 135 (2011),  doi:10.1016/j.physletb.2011.07.058,  arXiv:1105.5824 [hep-ph].


\bibitem{d0-polemass-ref}  D0 Collab. (V. M. Abazov {\it et al}.), Phys. Lett. B {\bf 703}, 422 (2011), doi:10.1016/j.physletb.2011.08.015, arXiv:1104.2887 [hep-ex].


\bibitem{cms-polemass-ref} CMS Collab., CMS Physics Analysis Summary CMS-PAS-TOP-11-008 (2011), https://cds.cern.ch/record/1387001.

\bibitem{atlas-polemass-ref} ATLAS Collab.,  Conference note ATLAS-CONF-2011-054 (2011), https://cds.cern.ch/record/1342551.

\bibitem{lagen-xsec} U. Lagenfeld, S. Moch and P. Uwer, Phys. Rev. D 80 (2009) 054009, doi:10.1103/PhysRevD.80.054009, arXiv:0906.5273 [hep-ph].

\bibitem{kido-xsec} N. Kidonakis, Phys. Rev. D 82 (2010) 114030, doi:10.1103/PhysRevD.82.114030, arXiv:1009.4935 [hep-ph].

\bibitem{ahrens-xsec} V. Ahrens, A. Ferroglia, M. Neubert, B. D. Pecjak and L. L. Yang, J. High Energy Phys. 09 (2010) 097, doi:10.1007/JHEP09(2010)097,  arXiv:1003.5827 [hep-ph].

\end{thebibliography}
\end{document}